\documentclass[reprint,prd]{revtex4-1}
\usepackage{natbib}
\usepackage{graphicx}
\usepackage[export]{adjustbox}
\usepackage{xcolor}
\usepackage{amssymb,amsmath}
\RequirePackage[colorlinks=true
,urlcolor=blue
,anchorcolor=blue
,citecolor=blue
,filecolor=blue
,linkcolor=blue
,menucolor=blue
,pagecolor=blue
,linktocpage=true
,pdfproducer=medialab
,pdfa=true
]{hyperref}

\begin{document}
\title{A Foreground Model Independent Estimation of Joint Posterior of CMB E mode Polarization over  Large Angular Scales}

\author{Ujjal Purkayastha$^1$, Vipin Sudevan$^1$, Rajib Saha$^1$}
\affiliation{$^1$Physics Department, Indian Institute of Science
	Education and Research Bhopal, \\ Bhopal, M.P, 462066, India.}

\begin{abstract}
       Ever since Cosmic Microwave Background (CMB) signal is being measured by various satellites 
       based observations with increasing experimental accuracies there has been a parallel increase 
       in the demand for  a CMB reconstruction technique which can provide accurate estimates of CMB signal and the theoretical 
       angular power spectrum along with reliable statistical error estimates associated with them. In this 
       work, we estimate the joint posterior of CMB E mode signal ({\bf S}) and corresponding theoretical angular power spectrum 
       ($C^E_{\ell}$) over large angular scales given the simulated polarization observations of future generation COrE 
       satellite mission. To generate samples from the joint distribution we employ the internal-linear-combination
       (ILC) technique with prior information of CMB E mode covariance matrix  augmented by a Gibbs 
       sampling technique of Sudevan and Saha 2020. We estimate the marginalized densities of ${\bf S}$
        and $C^E_{\ell}$ using the samples from full-posterior. The best fit cleaned E mode map and the 
       corresponding angular power spectrum agree well with the input E mode map and the sky power spectrum
       implying accurate reconstruction using COrE like observations.  
       Using the samples 
       $C^E_\ell$ of all Gibbs chains we estimate 
       the likelihood function $P(C^E_{\ell}|{\bf D}$) of any arbitrary $C^E_\ell$  given  simulated observed maps 
       (data, {\bf D}) of COrE mission following Blackwell-Rao estimator. The likelihood function can be seamlessly integrated 
       to the cosmological parameter estimation method.  Apart from producing an accurate estimate of E mode signal
       over large angular scales our method also builds  a connection between the component reconstruction and 
       reliable cosmological parameter estimation using CMB E mode observations over large angular scales. 
       The entire method does not  assume any explicit models for E mode foreground components 
       in order to remove them, which is an attractive property since foreground modelling uncertainty 
       does not pose as a challenge in this case.
       
\end{abstract}


	\maketitle 
	
	\section{Introduction}
	\label{Intro}

The weak E mode polarization signal of Cosmic Microwave Background (CMB) radiation~\cite{Penzias1965} 
anisotropies  were generated in early Universe by Thomson scattering at the surface of last scattering 
and by  the scattering of the CMB photons by the 
ionized Hydrogen and Helium atoms in the intergalactic medium  during reionization epoch~\cite{2010ASPC..438..276C} 
in relatively recent past. CMB E mode signal provides valuable insights into the era of reionization by 
breaking degeneracies between scalar field fluctuation amplitude and optical depth of
the reionization era. The reionization bump over large angular scales of CMB E mode 
signal encodes unique information about the reionization optical depth~\cite{Hiramatsu:2018nfa}. The shape and location
of this peak is sensitive to the epoch of reionization~\cite{PhysRevLett.91.241301,1997PhRvD..55.1830Z}. For accurate estimation 
of cosmological parameters from the CMB E mode signal over the  large angular scales a foreground 
removal technique that can provide accurate  estimates of the signal 
along with the statistical error estimates is necessary to be employed.

An important derivable in context of CMB  analysis is the joint posterior $P({\bf S}, C^E_\ell |{\bf D})$ 
of the CMB E mode signal ${\bf S}$ and  theoretical angular power spectrum $C^E_\ell$ given the  observed (E mode) maps, ${\bf D}$ since 
following Bayesian concept all the information that can be derived about ${\bf S}$ and $C^E_\ell$ from  ${\bf D}$
is encoded in this function. The posterior density can be used to best-fit values of  the random 
variables ${\bf S}$ and $C^E_\ell$ along with the associated error-intervals. The posterior can be marginalized over the E mode 
signal to estimate the likelihood function $P(C^E_\ell | {\bf D})$ which  
plays a central role in estimating the cosmological parameters.  The posterior and hence the likelihood function 
derived from the cleaned CMB E mode signal are however expected to be different for different CMB 
component reconstruction method since each is characterized effectively by a different foreground removal filter. 
Hence these functions need to be accurately determined for a CMB 
reconstruction technique for accurate estimation and correct interpretation of cosmological parameters.  In this article, we estimate both the
joint posterior density and likelihood function of the cleaned E mode maps over large angular scales using 
internal linear combination (ILC) method along with the Gibbs sampling technique~\cite{Gibbs1984}. Since the likelihood function 
estimated by us is calculable for any choice of $C^E_\ell$ it can be readily used to calculate cosmological parameters 
integrating with a Markov Chain Monte Carlo (MCMC) method~\cite{Metropolis1953}.

A major step towards estimating the joint conditional distribution and the likelihood function 
discussed above is process of reconstruction of  foreground minimized CMB E maps. 
Two major polarization 
foreground components are synchrotron at low frequency and thermal dust at high frequency. These contamination are very strong at large angular scales and can
almost cover the reionization bump~\citep{2003NewAR..47.1127B}. Reconstruction of  weak E mode background 
CMB signal by removing the strong foregrounds therefore becomes a challenging task.  
The method may be further limited by the presence of substantial amount of detector noise in the observed maps. 
Fortunately, many future generation sensitive CMB experiments are being  designed to accurately measure 
the weak CMB polarization anisotropy with sufficiently large signal to noise ratio. 

Two distinct routes have been followed in the literature in order to accurately 
reconstruct the CMB component by removing the foregrounds. The first type of them minimizes
contributions from all non-CMB components thereby leaving only a CMB signal, without, however, using 
any explicit foreground models. The other type of the methods  requires prior knowledge about the 
frequency dependence and or morphological pattern of the different foreground components present in the observed CMB
maps. Such methods are known as foreground model-dependent methods like Wiener Filtering~\cite{1994ApJ...432L..75B,1999MNRAS.302..663B}, Gibbs sampling
approach~\cite{Groeneboom:2008fz,Eriksen:2007mx,Eriksen:2007mp,Wandelt:2003uk}, template fitting method~\cite{FernandezCobos:2011bm}, 
the maximum entropy method~\cite{Gold:2008kp}, a MCMC~\citep{Gold2009} method. 
This method works based on principle of inclusiveness in which one exploits freedom 
of large scale modelling of each and every physical component present in the sky. 
The first type of method, on the contrary, requires very little assumption about the 
foreground components which are to be removed and as far as CMB reconstruction alone 
is concerned as is the case for cosmological analysis, becomes a much simpler problem 
to focus. Various model-independent methods have been developed such as Independent Component Analysis 
(ICA)~\cite{2003MNRAS.344..544M,2002MNRAS.334...53M,2010MNRAS.402..207B}, 
ILC~\cite{Tegmark1996}, Correlated Component Analysis (CCA)~\cite{Bedini:2004dd}.

ILC is a foreground minimization method in which  in order to
obtain a cleaned CMB signal one makes a simple assumption that the foreground and noise spectra are different from the
CMB power spectrum, which follows a black-body spectrum~\cite{Mather:1993ij}. The cleaned CMB map obtained using
the ILC method is not susceptible to inaccuracies in foreground  modeling. In the ILC method,
a cleaned CMB map is obtained by linearly combining multi-frequency observed foreground
contaminated CMB maps using some amplitude terms known as weight factors. These weights are subjected to the constrain 
that the sum of all weights should be unity. In the  ILC
method these weights are estimated analytically by performing a constrained minimization of
the variance of the cleaned CMB map. The analytical nature of the weights is an added 
advantage of the method since it avoids issues related to numerical convergences that may 
be present in numerical approximate methods. 

{\it Since we use ILC method for E mode CMB reconstruction in the current article the posterior density and 
the likelihood function estimated by us become independent on the explicit modelling of polarized 
foregrounds. Modelling uncertainty for the polarized foregrounds does not become any issue for these
estimates. Moreover, as we demonstrate in this work  and since we use a large number of frequency maps in our analysis 
the effects of residual foregrounds in the cleaned E modes map and angular power spectrum are 
negligible. The posterior density and likelihood function derived in this work therefore can 
safely be assumed to be free from effects of such foreground residuals.} For estimation of the joint
conditional posterior using explicit models of foreground components we refer to~~\cite{Eriksen2008}.

The ILC method has been studied extensively in~\cite{Tegmark1996,Tegmark2003,Saha2006,Eriksen2008,
  Saha2008,Leach2008,Kim2008,Samal2010,Delabrouille2009,10.1111/j.1365-2966.2011.19770.x}. 
  A global ILC method was proposed~\cite{2018ApJ...867...74S} where the weights are estimated by
 minimizing a CMB covariance weighted variance instead of the usual variance in the cleaned maps. 
  The authors~\cite{Purkayastha:2020snm} extended this method to obtain a cleaned CMB E mode map at
 large angular scales. Recently~\cite{2020arXiv200102849S} a detailed analysis of the impact of
	random residual calibration errors present in the observed CMB temperature  maps 
have been studied for the case of~\cite{2018arXiv181008872S}  and was shown that the effects 
of these random calibration errors are small.

We have organized our paper as follows. In Section~\ref{formalism} we review the basic formalism of our
 method. In Section~\ref{input maps} we discuss the input maps that we have used in
the present analysis and the methodology in Section~\ref{method}. We present the results of
our analysis in Section~\ref{results}. In Section~\ref{br} we discuss the Blackwell-Rao
estimator for estimating continuous likelihood distribution of the CMB theoretical E mode
angular power spectrum. Finally we check for the  convergence of the Gibbs chains in
Section~\ref{convergence} and present the discussion and conclusion of our work in  
Section~\ref{disc}.
	
	\section{Basic Formalism}
	\label{formalism}

In this section we briefly review the basic formalism  of our method of estimating joint 
conditional density of CMB E mode signal. The method consists of two interweived steps 
of estimating the cleaned E mode CMB map by employing a modified version of ILC 
algorithm~\cite{2018ApJ...867...74S,Purkayastha:2020snm} and sampling the E mode theoretical angular power spectrum from 
the respective conditional density~\cite{2018arXiv181008872S}. Following Gibbs theorem after burn in rejection the 
collection of samples of cleaned E mode map and angular power spectrum forms samples 
from the desired joint posterior density.

The observables for the linear polarization of CMB are described by the Stokes Q and U 
parameters. Linear combinations of these variables  under a local coordinate transformation 
transform as spin $\pm 2$ quantities~\cite{1997PhRvD..55.1830Z,1997ApJ...488....1Z}. 
These spin functions can be expanded in terms of the spin $\pm 2$ basis functions 
${}_ {\pm 2} Y _ {\ell,m}(p)$ following, 
\begin{equation}
Q(p)\pm i U(p) = {\sum_{\ell m}} {}_{\pm 2} a _ {\ell m}{} _ {\pm 2} Y _ {\ell m}(p) \, ,
\end{equation}
where $p$ denotes the pixel index and ${}_{\pm 2} a _ {\ell m}$ represents the $\pm 2$
spherical harmonic coefficients. One can form linear combinations of the spin 
harmonic coefficients to obtain
\begin{equation}
a_{\ell,m}^{E} = -\frac{1}{2}({} _ {+ 2} a _ {\ell m}{} + {} _ { -2} a _ {\ell m}{}) \, ,
\label{emode}
\end{equation}
using which spin-0 CMB E mode map can be defined as follows,
\begin{equation}
E(p) = \sum_{\ell m}^{}a_{\ell m}^{E}Y_{\ell m}(p) \, ,
\label{emode1}
\end{equation}
where $Y_{\ell m}(p)$ represents the spin-0 spherical harmonics.

In this article we work directly in the E mode basis instead of Q, U basis from which 
E mode angular power spectrum $\hat C^E_\ell$ can be obtained by a simple spin-0 spherical 
harmonic transformation. Using E mode maps in 
the analysis helps since it avoids additional spin 2 spherical harmonic transformations necessary 
to obtain the E mode angular power spectrum at each Gibbs iteration while reducing the total disk 
storage requirement by half. The E mode maps can of course be converted to equivalent Q,U maps in
 a lossless fashion if desired.  Since we use  full-sky Q, U maps for converting to E mode maps 
the problem of leakage between E and B mode signal~\cite{Challinor:2004pr} does not arise.

Let us assume that we have observations of CMB E mode signal ${\bf S}$ at $n$ different frequencies. 
The observed map ${\bf d}_i$ at frequency $\nu_i$ is given by, 
\begin{equation}
{\bf d}_i = {\bf S} + {\bf F}_i + {\bf n}_i \, ,
\label{di} 
\end{equation}
where  ${\bf F}_{i}$ denotes net E mode polarization from all the foreground components at the 
the $i^{\tt th}$ frequency  and  ${\bf n}_{i}$ represents the detector noise contamination.
We assume each of the input frequency maps are already smoothed by a common beam and pixel
window functions so that we can safely remove any explicit reference to these smoothing 
effects in Eqn.~\ref{di}. 
Each of the bold-faced variables in Eqn.~\ref{di}  represents a column vector of size $N_{pix}$, total number 
of pixels in the observed maps. The observed data ${\bf D} = \{{\bf d}_1, {\bf d}_2,...,{\bf d}_n \})$
is represented by a matrix of size $N_{pix} \times n$. 

To map out the posterior density $ P({\bf S}, C^E_\ell \vert {\bf D})$ one approach is to draw 
large number samples from it without direct evaluation of the function. An alternative method 
to achieve this is to employ Gibbs sampling in  which samples are drawn from the two conditional
densities which are easier to sample. In particular the $(i+1)^{th}$ Gibbs CMB E mode signal is obtained 
by drawing sample from
\begin{equation}
{\bf S}^{i+1} \leftarrow P_1({\bf S}|{\bf D},C_\ell^{E,i}) \, ,
\label{cmbsamp}
\end{equation} 
where $C_\ell^{E,i}$ represents the corresponding theoretical angular power spectrum sampled at the previous 
step. $C_\ell^{E,i+1}$ is obtained by sampling the conditional density  
\begin{equation}
C_\ell^{E,i+1} \leftarrow P_2(C^{E,i}_\ell|{\bf D},{\bf S}^{i+1}) \, .
\label{clsamp}
\end{equation}
Using the pair of samples at the $(i+1)^{th}$ iteration  we repeat the two sampling steps 
(Eqns.~\ref{cmbsamp} and~\ref{clsamp}),  for a large number of iterations  until convergence is achieved.
Ignoring the initial  burn-in phase the samples drawn from the two conditional densities are 
then samples from the joint posterior density.

How to draw a samples of CMB E mode signal given ${\bf D}$ and $C^{E}_\ell$ ? This is achieved 
by estimating the cleaned E mode CMB signal given the data and a theoretical CMB E mode 
angular power spectrum following  the foreground removal method described 
by~\cite{2018ApJ...867...74S,Purkayastha:2020snm}. The cleaned CMB E mode map $\hat{\bf S}$ which is an estimator of 
the true CMB E mode {\bf S} signal is obtained following the ILC method, 
\begin{equation}
\hat{\bf S} = \sum_{i=1}^{n} w_i {\bf d}_i \, , 
\label{cmbmap}
\end{equation}
where,
$w_i$ represents  the weight corresponding to the $i^{\tt th}$ frequency channel.
Since ${\bf S}$ is independent of frequency due to the black-body nature of its frequency 
 spectrum the weights  follow a constraint that they  sum to unity, i.e
$\sum_{i=1}^{n} w_i = 1$ in order that the cleaned map $\hat {\bf S}$  has no normalization bias 
as far as the underlying CMB E mode signal ${\bf S}$ is concerned. Using this condition on weights, we perform a constrained
minimization of the CMB E mode covariance weighted variance, $\sigma^2$ defined as  
\begin{equation}
\sigma^2 = {\bf S}^{T}{\bf C}^{\dagger}{\bf S} \, ,
\end{equation}
to estimate the weights. {\bf C} represents the CMB E mode theoretical covariance matrix and ${}^\dagger$ denotes the Moore-
Penrose generalized inverse~\cite{1955PCPS...51..406P}. Using a Lagrange's multiplier approach, we obtain the
choice of weights which minimizes $\sigma^2$ as
\begin{equation}
{\bf W} = \frac{{\bf e}^T {\bf A}^{\dagger}}{{\bf e}^T {\bf A}^{\dagger}{\bf e}} \, , 
\label{weight}
\end{equation}
where {\bf W} is a $(1 \times n)$ weight vector with elements {\bf W} = $\{ w_1, w_2,..
,w_n\}$ and the $(i,j)^{\tt th}$ element of the {\bf A} matrix is in pixel space is
given by,
\begin{equation}
{A}_{ij} = {\bf d}^T_i {\bf C}^\dagger {\bf d}_j \, .
\label{aij}
\end{equation}
Since ${\bf C}$ is a dense $N_{pix} \times N_{pix}$ matrix computing 
 Eqn.~\ref{aij} at every  Gibbs iteration becomes computationally very expensive. 
It is greatly advantageous to work in the harmonic space where Eqn~\ref{aij} is simpler 
to compute,
\begin{equation}
A_{ij} =  \sum_{\ell=2}^{\ell_{max}}(2\ell+1)\frac{\sigma^{ij}_\ell}{C^{E\prime}_{\ell}} \, , 
\label{amat}
\end{equation}
where $\ell_{max}$ represents the maximum multipole used in the analysis. 
$\sigma^{ij}_\ell$ denotes the cross angular power spectrum  between ${\bf d}_i$ and ${\bf d}_j$ and 
$C_\ell^{E\prime}$ represents  the beam and pixel smoothed CMB  EE theoretical power spectrum i.e.,
\begin{equation}
{C_{\ell}^{E\prime}} = {C^E_{\ell}{B^2_{\ell}} {P^2_{\ell}} } \, ,
\end{equation}
where $C^E_\ell$  does not contain any smoothing effects and  $B_{\ell}$ and  $P_{\ell}$ are 
respectively the beam and pixel window functions.

To draw samples of $C^E_\ell$ given ${\bf S}$ and ${\bf D}$ we first 
obtain the conditional density on $P_2(C^E_\ell|\textbf{S},\textbf{D})$
in terms of the variable $y=\hat{C}^E_\ell (2\ell+1)/{C^E_\ell}$ as~\cite{2018arXiv181008872S}
\begin{eqnarray}
P_{2}(C^E_{\ell}|\hat{C}^E_\ell) \propto y^{-(2\ell-1)/2-1}  \text{exp}\left[-\frac{y}{2} \right] \, ,
\label{eq_y}
\end{eqnarray}
where, $\hat{C}^E_\ell$ is estimated from the cleaned CMB E mode map and 
a uniform prior on $C^E_\ell$ is assumed. Eqn.~\ref{eq_y} shows
that the variable  $y$ follows a $\chi ^2$ distribution  with $ 2\ell-1$ degrees of
freedom. Therefore, in order to sample a CMB theoretical power spectrum, we draw
$y$ from the $\chi^2$ distribution of $2\ell-1$  degrees of freedom and then find
$C^E_\ell$ as  follows
\begin{equation}
C^E_\ell = \hat{C}^E_\ell (2\ell + 1)/y.
\end{equation}

\section{Input Maps}
	\label{input maps}
	In our analysis, we simulate foreground and noise contaminated CMB E mode maps at the
	 COrE  frequency channels ranging from $60$ GHz to $340$ GHz. The higher frequencies are 
         avoided as they have comparatively large level of detector noise contamination. The frequency maps 
        used in this work along with the beam widths and noise levels are listed in Table~\ref{simmaps}. 
	
	\subsection{CMB Maps}
	We generate  random realizations of the CMB Stokes Q and U   maps using the {\tt anafast} 
        compatible to Planck theoretical CMB angular power 
        spectrum~\cite{2016A&A...594A..13P} at HEALPix (Hierarchical Equal Area IsoLatitude 
        Pixellation of sky)~\cite{2005ApJ...622..759G} resolution 
        $N_{\tt side} =16$. We then smooth the resulting map using a Gaussian beam window  
        of FWHM $9^{\circ}$ and convert it to the CMB E mode map using the  {\tt synfast} facility.
	
	\subsection{Foreground model}
\label{fg_nse_input}

\begin{figure}
 \includegraphics[scale=0.275]{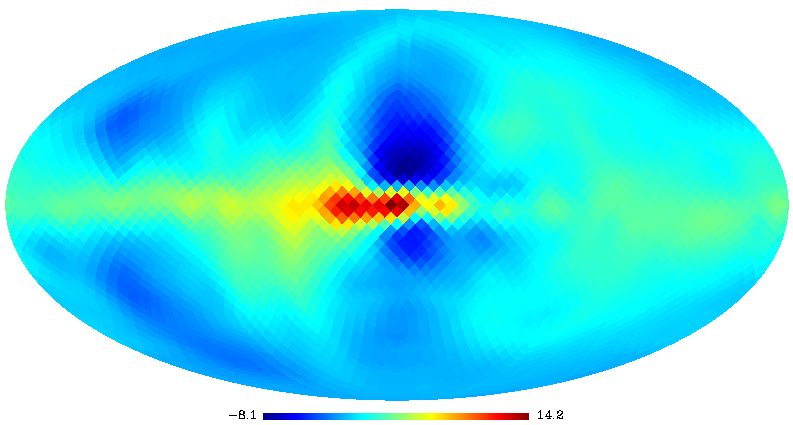}
 \includegraphics[scale=0.275]{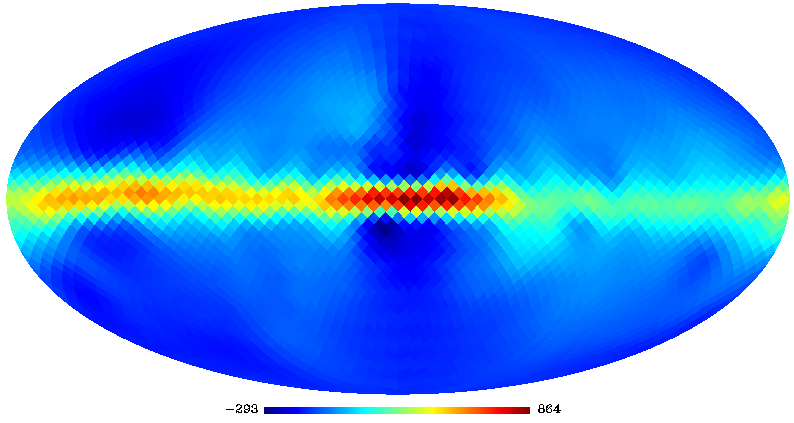}
 \caption{In the top  panel we show the synchrotron E mode emission map for 60 GHz, which is a 
 major foreground in lower frequencies ($<$ 100 GHz). The bottom panel shows  the thermal 
 dust E mode map for $340$ GHz. The thermal dust emission is a major source of foreground 
 at high frequency channels. Clearly we can see that the morphological pattern of the two  
foregrounds  are different from each other. Strong emission both in the positive  and  
 negative pixel values  exist near the galactic plane for synchrotron and thermal dust components.}
 \label{synch_th}
\end{figure}

\begin{figure*}
\centering
\includegraphics[scale=1.3]{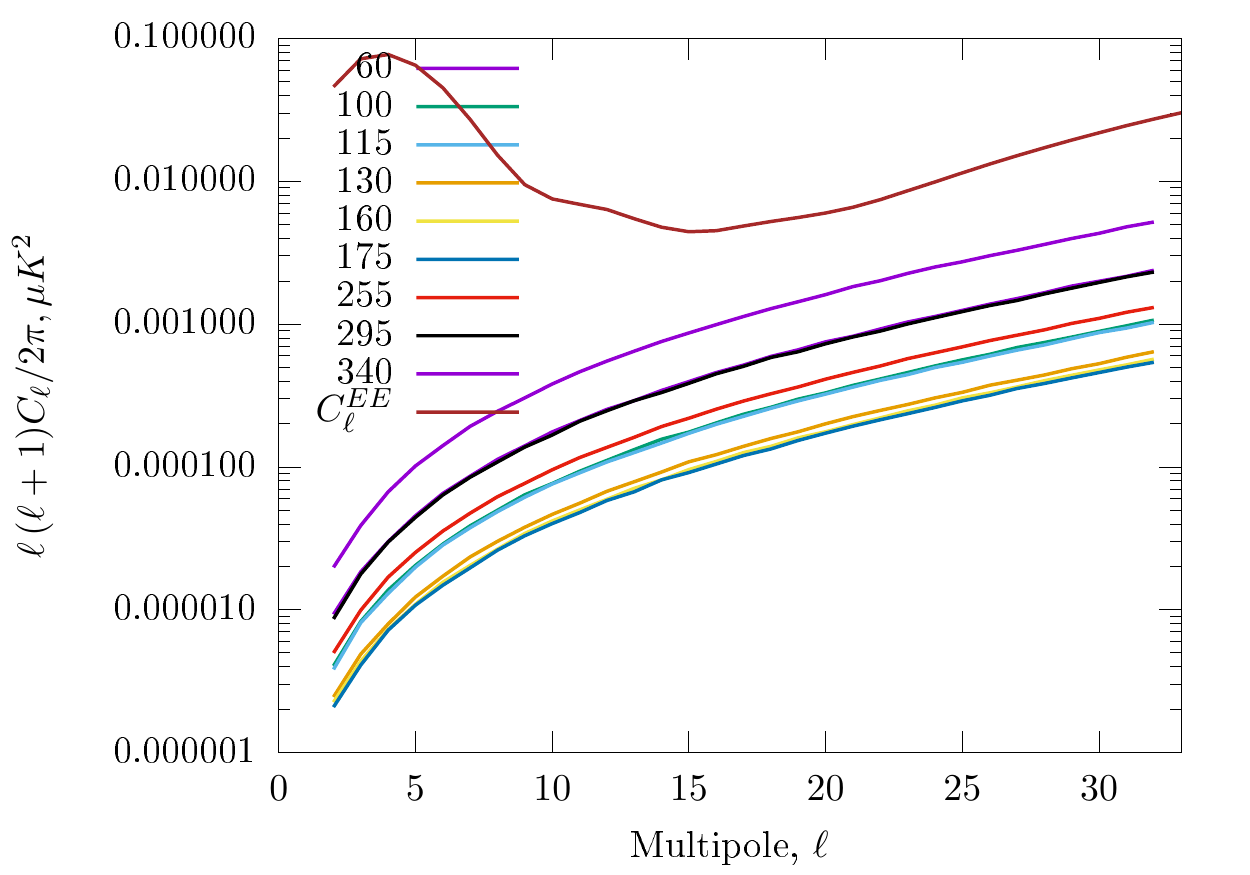}
\caption{We show the  mean noise E-mode power spectra obtained from 100 simulations
                        for some of the COrE frequencies. All the $C_\ell ^E$ have been deconvolved
                        by the beam and appropriate pixel window function. The theoretical CMB EE   
                        power spectrum is plotted as brown line  for reference. From the
figure we see that the the CMB theoretical angular power spectrum is well above the noise power.}
\label{noisecls}
\end{figure*}

        Synchrotron and thermal dust are two major  foreground  components that contaminate CMB
polarization signal. We first generate Stokes Q and U maps for synchrotron and thermal dust components
at the $15$  COrE frequencies  between $60$ to $340$ GHz  following a procedure similar to~\cite{Rema2018}. To generate the
synchrotron Stokes maps we use WMAP 23 GHz (K band) Stokes maps smoothed by a effective Gaussian beam
window function of FWHM $1^{\circ}$ at  $N_{\textrm {side}} = 512$.  In thermodynamic $\mu K$ temperature unit
the synchrotron Stokes maps $Q_s(\nu, p)$, $U_s(\nu, p)$ at a  frequency $\nu$ (in GHz) are  given by,
\begin{eqnarray}
\begin{split}
Q_s(\nu, p) = 1000\left(\frac{a(\nu)}{a(\nu_s)}\right)Q_s(p) \left(\frac{\nu}{\nu_s}\right)^{\beta_s}\,, \\
U_s(\nu, p) = 1000\left(\frac{a(\nu)}{a(\nu_s)}\right)U_s(p) \left(\frac{\nu}{\nu_s}\right)^{\beta_s}\,,
\end{split}
\label{Synch}
\end{eqnarray}      
where $p$ denotes the pixel index,  $Q_s(p)$ or $U_s(p)$  represents the  synchrotron  Q or U map at
the reference frequency $\nu_s =  23$ GHz, as observed by WMAP {in mk thermodynamic temperature unit}, $a(\nu)$ represents the
antenna to thermodynamic conversion factor and $\beta_s  = -3.0$ which represents the pixel independent
synchrotron spectral index. As mentioned in~\cite{Rema2018} the choice of such a spectral index
corresponds to the typical mean synchrotron spectral index measured at the CMB observation window.

We generate thermal dust polarization maps at the five COrE frequencies by extrapolating Planck $353$ GHz
thermal dust optical depth map  as produced by generalized needlet ILC (GNILC)  algorithm~\citep{ThDustGNILC2016}.
Using the GNILC  map is advantageous since it provides a significantly improved
picture of galactic thermal dust emission by separating the cosmic infrared background (CIB)
anisotropies. We downgrade the $353$ GHz GNILC optical depth map at $N_{\textrm {side}} = 512$ and
smooth it by a Gaussian beam function of FWHM = $\sqrt{(60^{\prime})^2 - (5^{\prime})^2}$ to bring the resulting map to
the effective beam resolution of $1^{\circ}$. We represent this map by $\tau_{353}(p)$. The thermal
dust intensity map $I(\nu,p)$ is then obtained following,
\begin{eqnarray}
I(\nu,p) = 10^{20}B(\nu, T_d) \tau_{353}(p) \left(\frac{\nu}{\nu_d} \right)^{\beta_d}\, ,
\label{Inu}
\end{eqnarray}  
where, $\beta_d = 1.6$  and $T_d = 19.4$K (e.g.,~\cite{Rema2018, ThDustGNILC2016}), $\nu_d = 353$ GHz and $B(\nu, T_d)$
represents the Planck function at a thermal dust temperature $T_d$,
\begin{eqnarray}
B(\nu, T_d) = \frac{2 h^3}{c^2}\frac{1}{\exp\left(\frac{h\nu}{k_BT_d}\right)-1} \, .
\end{eqnarray}
Since $B(\nu, T_d)$ physically represents intensity, in metric system its unit is
$\textnormal{Joule sr}^{-1} {\textnormal m}^{-2} {\textnormal s}^{-1} {\textnormal Hz}^{-1}$. Therefore,
the numerical factor of $10^{20}$ in Eqn.~\ref{Inu} estimates
dust intensity in $\textnormal{MJy sr}^{-1}$ unit.  
We estimate thermal dust Stokes Q, U  polarization maps using the intensity map
of Eqn.~\ref{Inu} following~\citep{PSM2013,Rema2018},
\begin{eqnarray}
\begin{split}
Q(\nu,p) = f_dg_d(p) \cos(2\gamma_d(p))I(\nu,p)\, , \\
U(\nu,p) = f_dg_d(p) \sin(2\gamma_d(p))I(\nu,p)\, ,
\end{split}
\label{dust_qu}
\end{eqnarray}
where $f_d$, $g_d(p)$ and $\gamma_d(p)$ represent respectively   pixel independent intrinsic
dust polarization fraction and pixel dependent dust geometric depolarization and polarization
angle. As in \cite{PSM2013,Rema2018} we take $f_d = 0.15$. Following these literatures
the polarization angles are estimated using WMAP K band Stokes Q and U maps (after smoothing to
effective resolution of $3^{\circ}$) using,
\begin{eqnarray}
\gamma_d(p) = \frac{1}{2}\tan^{-1}\left(-U_{23}(p), Q_{23}(p)\right) \, .
\label{gammad}
\end{eqnarray}
The depolarization factor is computed from the properly scaled version of the degree of linear
polarization at $23$ GHz as obtained by extrapolating the $408$ MHz Haslam synchrotron
template (${ I_{0.408}}(p)$)  to  $23$ GHz assuming a constant spectral index $-3.0$. Prior
to using the Haslam template we subtract   a residual monopole offset of $8.33K$ in antenna temperature
unit and smooth the corrected map to an effective resolution of $3^{\circ}$ at $N_{\textnormal {side}} = 512$.  The
depolarization factor  is obtained following,  
\begin{eqnarray}
g_d(p) = \frac{\sqrt{Q^2_{23}(p) + U^2_{23}(p)}}{1000 f I_{0.408}(p)\left(23.0/0.408\right)^{-3}} \, ,
\label{depol}
\end{eqnarray}
where the factor $f$ in the denominator represents the product of synchrotron polarization
fraction ($0.75$) and antenna to thermodynamic conversion factor at $23$GHz. Once we have obtained
the polarization angles and depolarization factors respectively following Eqns.~\ref{gammad} and
\ref{depol} we use them in Eqn.~\ref{dust_qu} to estimate thermal dust Stokes parameters at COrE
frequencies.

 \begin{figure*}
 \centering
 \includegraphics[scale=0.33]{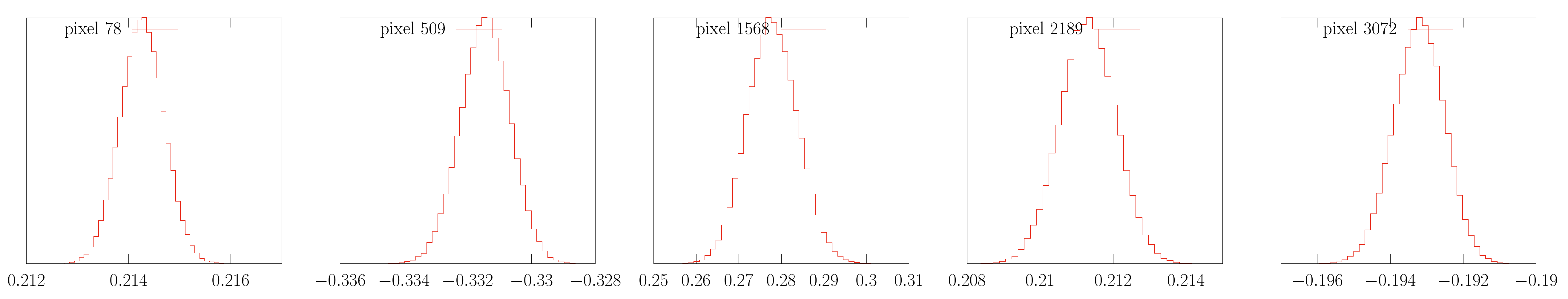}
 \caption{We show the normalized density plots for some of the pixels of the CMB E mode maps.
 The normalization is performed by dividing by the corresponding modal value such that the
 position of the peak  corresponds to unity in all the plots. We see that the densities are
 nearly symmetrical with minor asymmetries
 around the tail. The horizontal axis is in the unit of thermodynamic $\mu K$.  }
\label{normalpix}
\end{figure*}

Using the full sky Stokes Q and U maps for synchrotron and thermal dust components as  discussed above
we obtain full sky E mode maps for these components at all the frequencies of this work at $N_{side}=16$ 
as follows. First we generate $a^E_{\ell m}$ at each frequency 
upto the maximum multipole $\ell_{max} = 32$  following Eqn~\ref{emode} and smooth them
by polarized Gaussian beam window function to an effective resolution of  FWHM $9^\circ$ and appropriate pixel window 
function. The E mode maps for each foreground component for the $15$ input frequencies are then obtained 
using Eqn.~\ref{emode1}. In top panel of Fig~\ref{synch_th} we show the synchrotron E mode emission map at $60$ GHz.
In the bottom panel we show the  thermal dust map at $340$ GHz.

 \subsection{Detector Noise Model }
Since presence of detector noise in the observed CMB maps is
unavoidable,  
we  generate realistic random realizations  of detector noise maps at all $15$ COrE frequencies of this work
using the COrE $4$ year noise model given in the Table~\ref{simmaps} (e.g., see ~\cite{2018JCAP...04..014D}). Noise levels for Stokes Q and U for the 15 COrE frequencies in the unit of $\mu$K.arcmin (thermodynamic) temperature
are given in the third column of Table~\ref{simmaps}.
We assume that the noise is Gaussian, isotropic and is uncorrelated from pixel to pixel. We also make a
further assumption that the Q and U noise maps are pixel  uncorrelated.
Mathematically this means that Stokes Q and U noise maps at pixel index $p$ can be represented as
\begin{eqnarray}
\langle Q_{i}(p)Q_{i}(p')\rangle = \sigma^2_{Qi}\delta_{pp'}\, ,\\
\langle U_{i}(p)U_{i}(p')\rangle = \sigma^2_{Ui}\delta_{pp'}\, ,\\
\langle Q_{i}(p)U_{i}(p')\rangle = 0 \, .
\end{eqnarray}
where $\sigma^2_{Qi} $ and $\sigma^2_{Ui} $ are the pixel noise variances for Q and U maps at 
frequency $\nu_{i}$ respectively and are assumed to be identical. The noise variances are given by
\begin{eqnarray}
\sigma^2_{Qi} = \sigma^2_{Ui} = \left(\Delta Q^2_i a_r\right)^2/\left(\Delta \Omega\right) \, ,
\end{eqnarray}
where $a_r$ represents the conversion factor from arcmin to radian and $\Delta \Omega = 4\pi/N_{\textrm {pix}}$ represents area  
for each $N_{\textnormal {side}} = 16$ pixel in unit of steradian ($N_{pix} = 12N^2_{\textnormal {side}}$).
We first generate random Gaussian Stokes Q and
U noise maps at $15$ COrE frequencies at $N_{\tt side} = 16$ and smooth them by the ratio of
a polarized Gaussian beam of FWHM $9^{\circ}$ and the original polarized beam given in Table~\ref{simmaps}.
This operation is performed in order to bring all the frequency maps to same beam resolution.
Finally, we convert the noise Stokes  maps to full sky E noise maps for all frequencies. 

We show the  mean noise power spectrum obtained from $100$ Monte Carlo simulations at some of the COrE frequencies in the
Fig.~\ref{noisecls} along with the CMB theoretical EE spectrum $C_{\ell}^E$.
We see that the mean noise power  is
lower than the theoretical CMB EE power spectrum for all mutlipoles.
\begin{table}
\centering
\begin{tabular}{lll}
\hline
Frequency &  Beam FWHM & $\Delta Q=\Delta U$            \\
(GHz)     &   (arcmin)     &
($\mu$K.arcmin)           \\  
\hline
60   & 17.87 & 7.49 \\
70   & 15.39 & 7.07  \\
80   & 13.52 & 6.78 \\
90  &  12.08 & 5.16 \\
100 &  10.92 & 5.02 \\
115  & 9.56 &  4.95  \\  
130 &  8.51 &  3.89 \\
145 &  7.68  &  3.61 \\
160 &  7.01  &  3.68\\
175 &  6.45  & 3.61\\
195 &  5.84  &  3.46 \\
220 &  5.23  &  3.81 \\
255 &  4.57  &  5.58 \\
295 &  3.99  &  7.42 \\
340 &  3.49   &11.10  \\
\hline
\end{tabular}
\caption{ COrE frequency maps used in this work}
\label{simmaps}
\end{table}

	\section{Methodology}
	\label{method}

	We implement our method to estimate the joint posterior for CMB E mode given the
	foreground contaminated CMB E mode maps in a (foreground) model independent manner.
	We use the foreground and noise model provided by the COrE science team and simulate
	foreground and noise contaminated CMB E mode input maps at all 15 COrE frequency channels
	at a pixel resolution defined by the HEALPix pixel resolution parameter $N_{\tt side} = 16$
	and smoothed by a Gaussian beam of FWHM $9^{\circ}$. Our foreground model consists of
	synchrotron and thermal dust. The noise model used is the one provided by 4 year COrE mission.
	The random CMB E mode realization
	used in the simulated input frequency maps is generated
	using the CMB theoretical angular power spectrum consistent with Planck 2015 results~\cite{2016A&A...594A..13P}.
	The details of the simulated input maps are given in the  Table~\ref{simmaps}.
	After simulating the foreground and noise contaminated input CMB E mode maps we
	remove the monopole and dipole components from all the maps. In our analysis,
	in order to sample the joint posterior density of CMB $P({\bf S}, C_{\ell}^E \vert {\bf D})$
	our  algorithm has 10 independent chains and each chain consist of 10000
	Gibbs steps. The initial point of each chain is obtained by randomly generating
	an initial value of $C_\ell^E$ from a uniform distribution within $\pm 3\Delta C_\ell^E $
	around the Planck best fit theoretical power spectrum, where $\Delta C_\ell^E$ is
	the cosmic variance error.
	
	We sample a CMB theoretical EE angular power spectrum and a cleaned
	CMB E mode map given the input CMB maps following the procedure as outlined in the
	Section~\ref{formalism} at each Gibbs step for every chain. At any given instance,
	we use Eqn.~\ref{cmbmap} to sample {\bf S}. The weights to be used in this equation are 
        obtained by using Eqn.~\ref{weight} in terms of matrix ${\bf A}$.  The elements of ${\bf A}$  
        are computed following
	the  Eqn.~\ref{amat} using the last sampled $C_{\ell}^E$ and the corresponding sky $\hat C^E_\ell$. 
         After an initial burn-in phase each of the Gibbs chains stabilizes.  We remove a conservative choice of 
        $50$ samples for  the burn-in period. Hence
	for analysis of the results  we are left with $99500$ samples of $C_\ell^E$ and ${\bf S}$.
        {\it We  note that the joint posterior density CMB E mode signal over large 
        angular scales derived in this work is independent on the polarized foreground models since 
        we use ILC method to remove the foregrounds. This causes the posterior density to be insensitive 
        to  any uncertainty that may exists in any one of the polarized foregrounds.} 	
	
	\section{Results}
	\label{results}
	In this section we discuss the results obtained by using the sampled E mode cleaned maps and theoretical angular power spectrum
        after burn-in rejection. 
        we also show that the noise levels of COrE $15$ frequencies used in this work are sufficiently small and can be ignored for large angular 
        analysis of the 
        CMB  E mode signal using our method. 
	\subsection{Cleaned Maps}
	\label{cleanedmaps}
	\begin{figure}
		\centering
		\includegraphics[scale=0.32,angle=0]{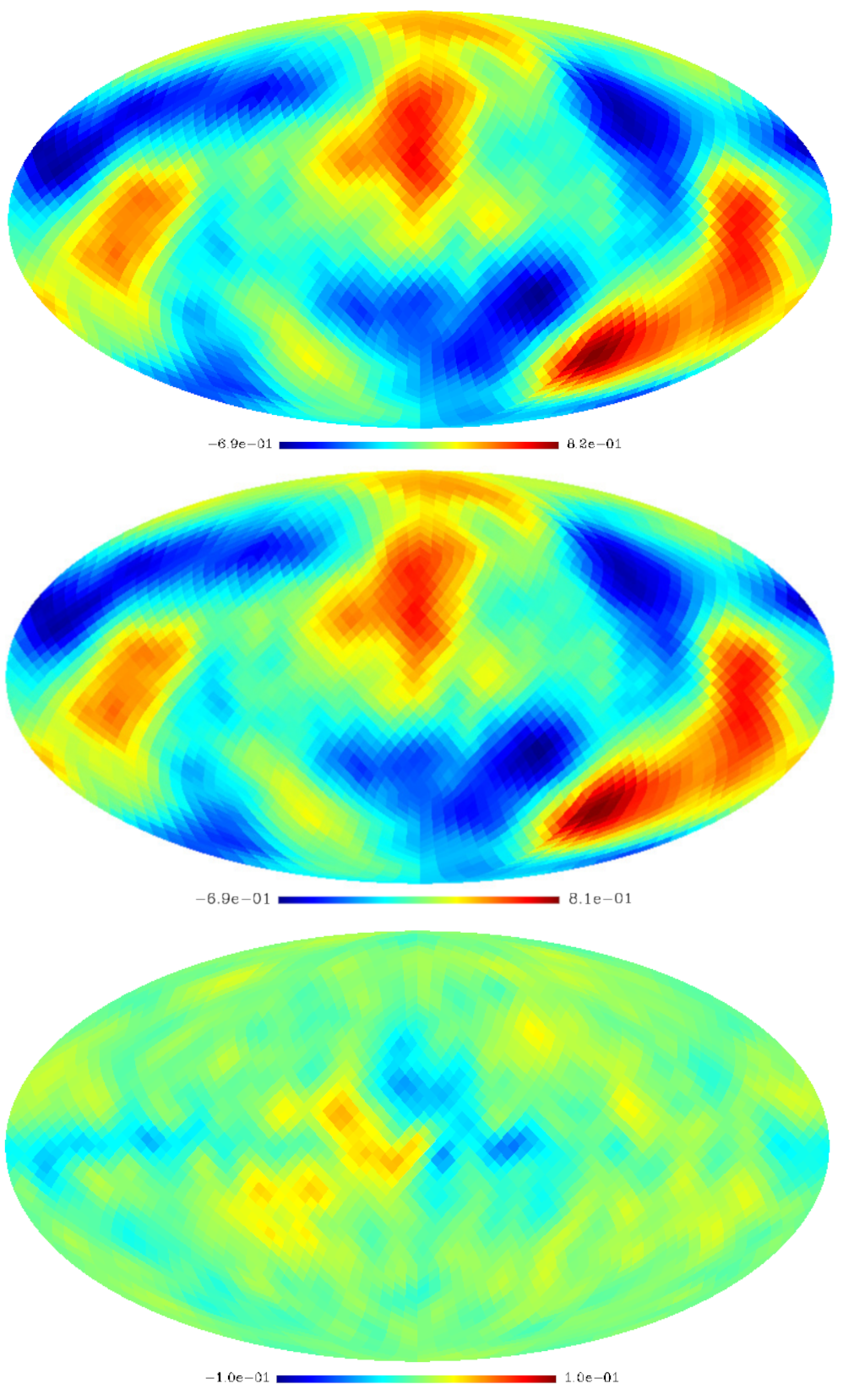}
		\caption{In the top panel we show the best fit cleaned CMB E mode map which is obtained by using the modal value of
			marginalized densities at each pixel. The middle panel shows the input CMB E mode map used in the simulations. 
                        The bottom panel shows the difference map obtained by subtracting the input map from the cleaned E mode 
			map. Both the top and bottom panel agree well with each other. } 
		\label{cleanedmapsfig}
	\end{figure}

       \begin{figure}
                \centering
                \includegraphics[scale=0.32,angle=0]{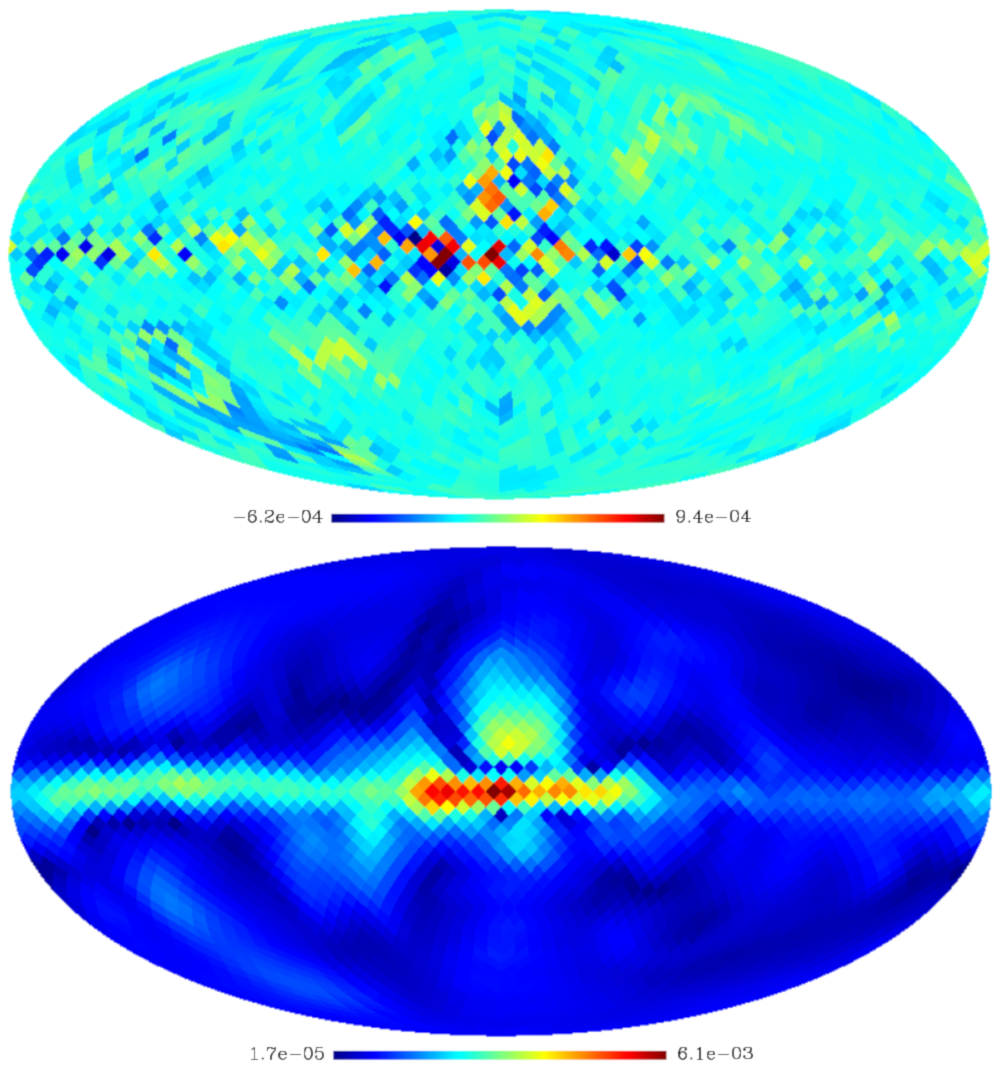}
                \caption{In the top panel we show  the  
                        difference between the the best fit cleaned CMB E mode map and the mean of all cleaned 
                  CMB E mode maps.  We see that our best fit cleaned map matches well 
                        with the mean cleaned CMB E mode map. In the bottom panel we show the standard deviation 
        map obtained from all the difference maps obtained by subtracting the input CMB map from the cleaned E mode maps. 
         We see  that the error while reconstructing
                        the pure CMB E mode signal using the input foreground and noise contaminated 
                         E mode maps using our method is very small.}
                \label{cleanedmapsfig1}
        \end{figure}

	\begin{figure*}
		\centering
		\includegraphics[scale=0.5]{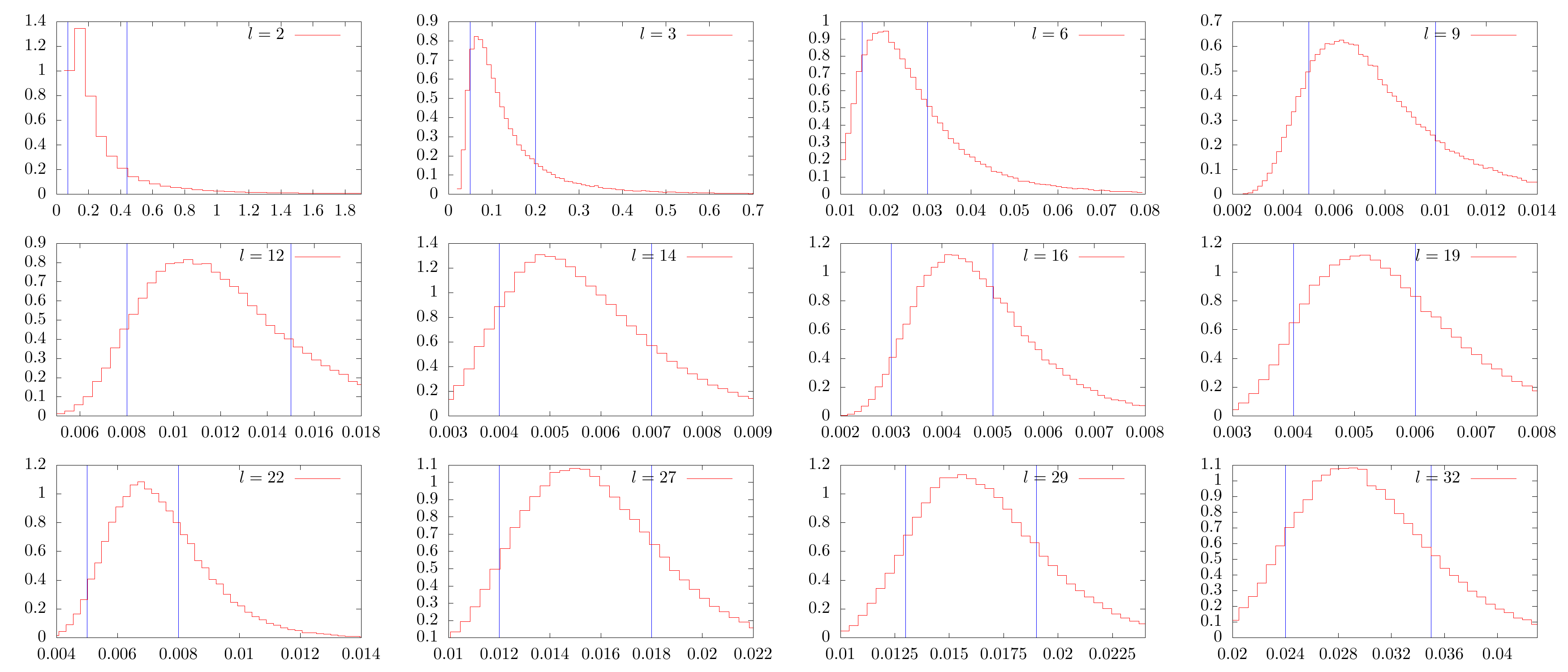}
		\caption{We show the normalized histogram plots of theoretical cleaned angular power spectrum of 
                        some of  multipoles whose horizontal
			axis represents the  $\ell(\ell +1)C_\ell^E/2\pi$ in units of $\mu K^2$. The region  
			between the sky blue vertical lines represents the $1-\sigma$ error interval around the best fit. The lower
			multipoles prominently show asymmetry for chi square distribution with low number of 
                        degrees of freedom which becomes gradually symmetrical as we move towards higher multipoles.}
		\label{normalcl}
	\end{figure*}
	
	We estimate the marginalized probability density of CMB E mode polarization signal at each pixel
	by using a total of $99500$ samples of the cleaned CMB E mode maps obtained from all the chains. 
        Each of these marginalized probability densities is converted to a normalized
	density corresponding to each pixel by dividing the respective marginalized probability
	densities with its mode value. We show these normalized probability densities corresponding to some randomly 
	chosen pixels in Fig.~\ref{normalpix}. We see from this figure that the normalized
	densities are very nearly symmetric with minor asymmetric tails. Finally, we estimate the
	best-fit cleaned CMB E mode map by assigning to each pixel the modal value of the corresponding
	histogram. We show our estimated best-fit cleaned CMB E mode map in the top panel of the
	Fig.~\ref{cleanedmapsfig}.  We see that the estimated posterior cleaned CMB E mode map do not
	contain any visible foreground residual contamination. The input CMB E mode map used in the simulation 
       is shown in the middle panel of this figure. The difference of the best-fit cleaned E mode map and the 
        input map is shown in the bottom panel of Fig.~\ref{cleanedmapsfig}. The best-fit map agree very well 
        with the pure CMB E mode map. We also estimate a mean cleaned CMB E mode
	map by taking the mean of all $99500$ samples of cleaned maps. We show the difference
	of the best-fit cleaned CMB E mode map and the  mean of all cleaned CMB E modes maps in the top panel  of the
	Fig.~\ref{cleanedmapsfig1}. We see that both the best-fit and the mean cleaned CMB E mode map matches
	with each other with a minor difference of the order of $\sim 10^{-4}$ at some pixels.
	In order to quantify the error while reconstructing a cleaned CMB E mode map at each iteration of
	the Gibbs sampling, from the foreground and noise contaminated CMB E mode maps we generate a
	standard deviation map using all the difference maps of cleaned CMB E mode maps and the input CMB E mode map. 
        The standard deviation map is shown 
	in the bottom panel of Fig.~\ref{cleanedmapsfig1}. Visually, we see that the reconstruction
	error is very small all over the sky. We see a marginal increase in the reconstruction
	error along the galactic plane particularly at the center owing to the strong
	foreground contamination in the input maps.
	
	\subsection{Cleaned Power Spectrum}
        \begin{figure*}
        \centering
        \includegraphics[scale=1.3]{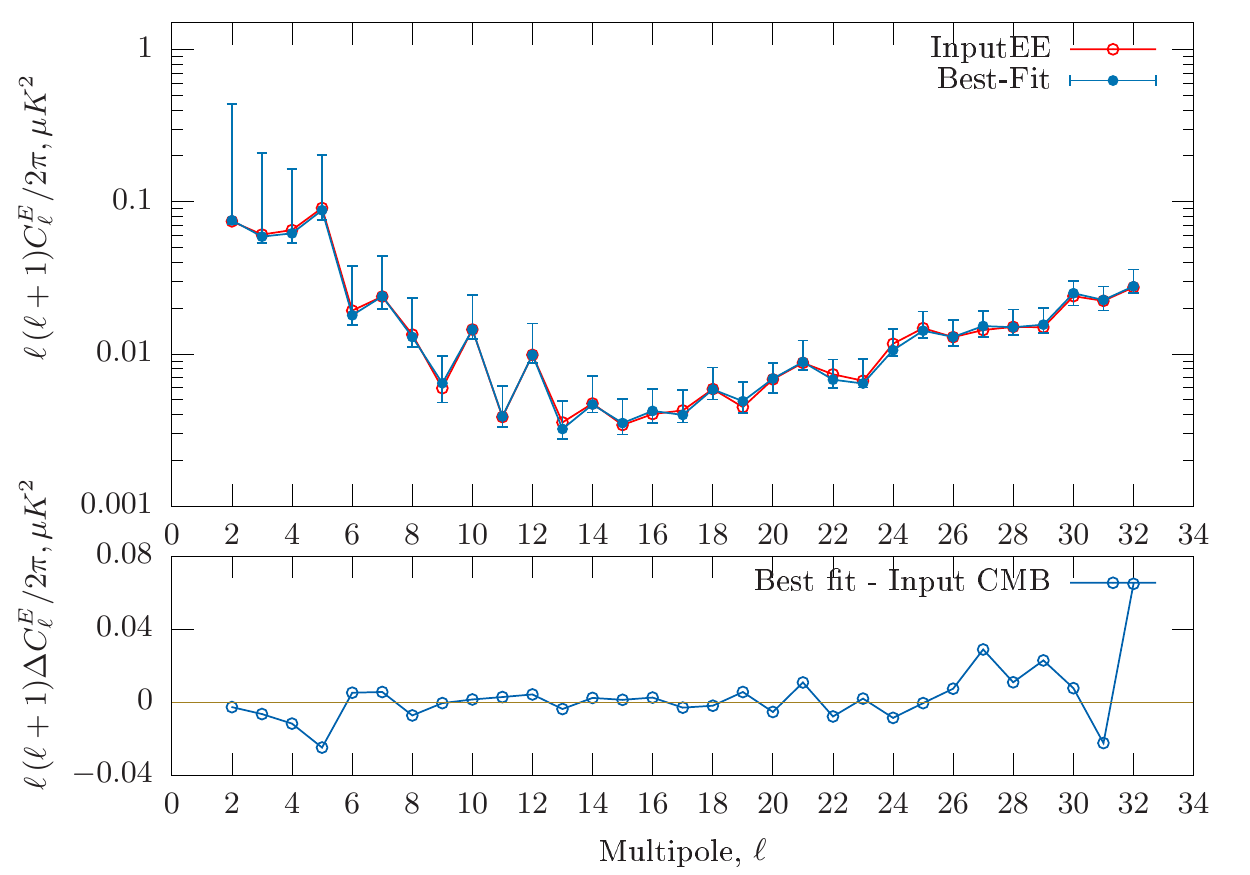}
        \caption{In the top panel we show the cleaned CMB EE power spectrum $ C_\ell^E $ with 
        asymmetric error bars in blue. The vertical axis is  in log scale $\ell(\ell +1)C_\ell^E/2\pi$ 
        in units of $\mu K^2$. Logarithmic axis is chosen to highlight the variation of the angular 
        power spectrum over a few  orders of magnitude. The power spectrum for the input CMB E mode 
        map is shown in red.  We show the difference between the green and  in the  bottom panel. 
        Both these spectra match well with each other.}
        \label{cleanedclfig}
        \end{figure*}

	We estimate the marginalized densities corresponding to mulitpoles of the CMB E mode theoretical
	angular power spectrum. In Fig.~\ref{normalcl}, we show the normalized densities for
	certain multipoles which are obtained in the same way as discussed in the
	Section~\ref{cleanedmaps}. In Fig.~\ref{normalcl}, the horizontal axis of each sub plot represents
	$\ell(\ell +1)C_\ell^E/2\pi$ in units of 
	$\mu K^2$. The blue vertical lines in each sub figure indicate the boundaries  
	for the asymmetric error bars with  $1 \sigma$ (68.27\%) confidence limits  for the
	corresponding sampled $C^E_\ell$. From the histogram plots it is evident that the lower  multipoles
	(e.g., $\ell = 2,3,..,6$) $C^E_\ell$ has  asymmetric distribution. The asymmetry gradually reduces
	as we go to higher multipoles ($\ell > 20$). In the top panel of Fig.~\ref{cleanedclfig},
	we show the best-fit CMB E mode theoretical angular power spectrum at all multipoles in  blue obtained by using the locations of the 
        peaks  of the marginalized angular power spectrum density function shown in Fig.~\ref{normalcl}. 
        Also plotted on the best-fit spectrum are the asymmetrical error bars corresponding to  $1 \sigma$ 
        confidence interval for each multipole. The error bars are large at low multipoles
	and gradually decreases as we go to higher multipoles. The red line indicates the the sky CMB 
        E mode angular power spectrum of this work. Both the spectra indicated by the blue and green match  
        with each other on the scales of this plot. 
	 In the bottom panel of Fig.~\ref{cleanedclfig}, we show
	the difference between best-fit CMB E mode theoretical angular power spectrum and input CMB  
        E mode $C_\ell^E$. We see that both the angular power spectra matches well with each other at all
	multipoles.
	
	\subsection{Residual Noise in the Cleaned E-mode CMB Maps}
	\begin{figure}
	\hspace{2.7cm}	\includegraphics[scale=0.54]{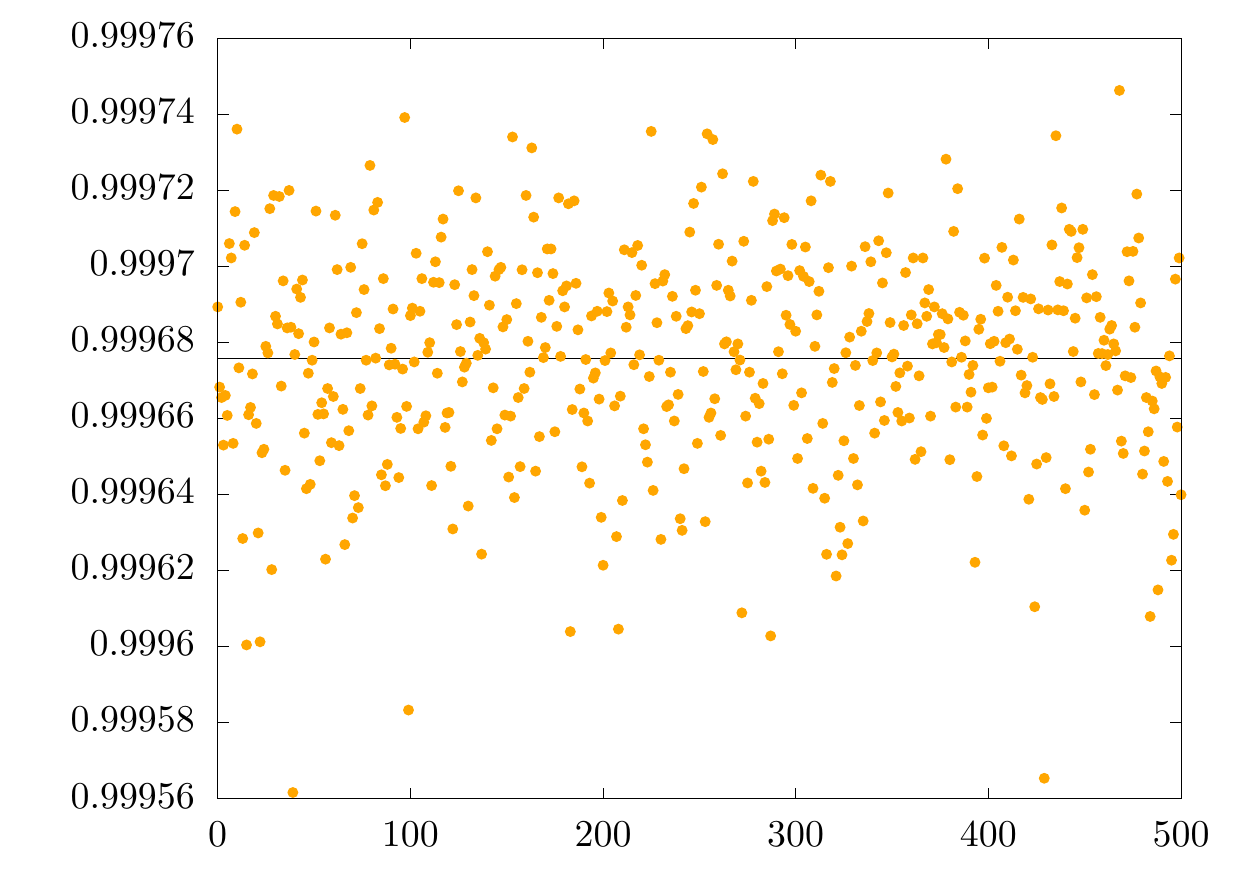}
		\includegraphics[scale=0.34,angle=90]{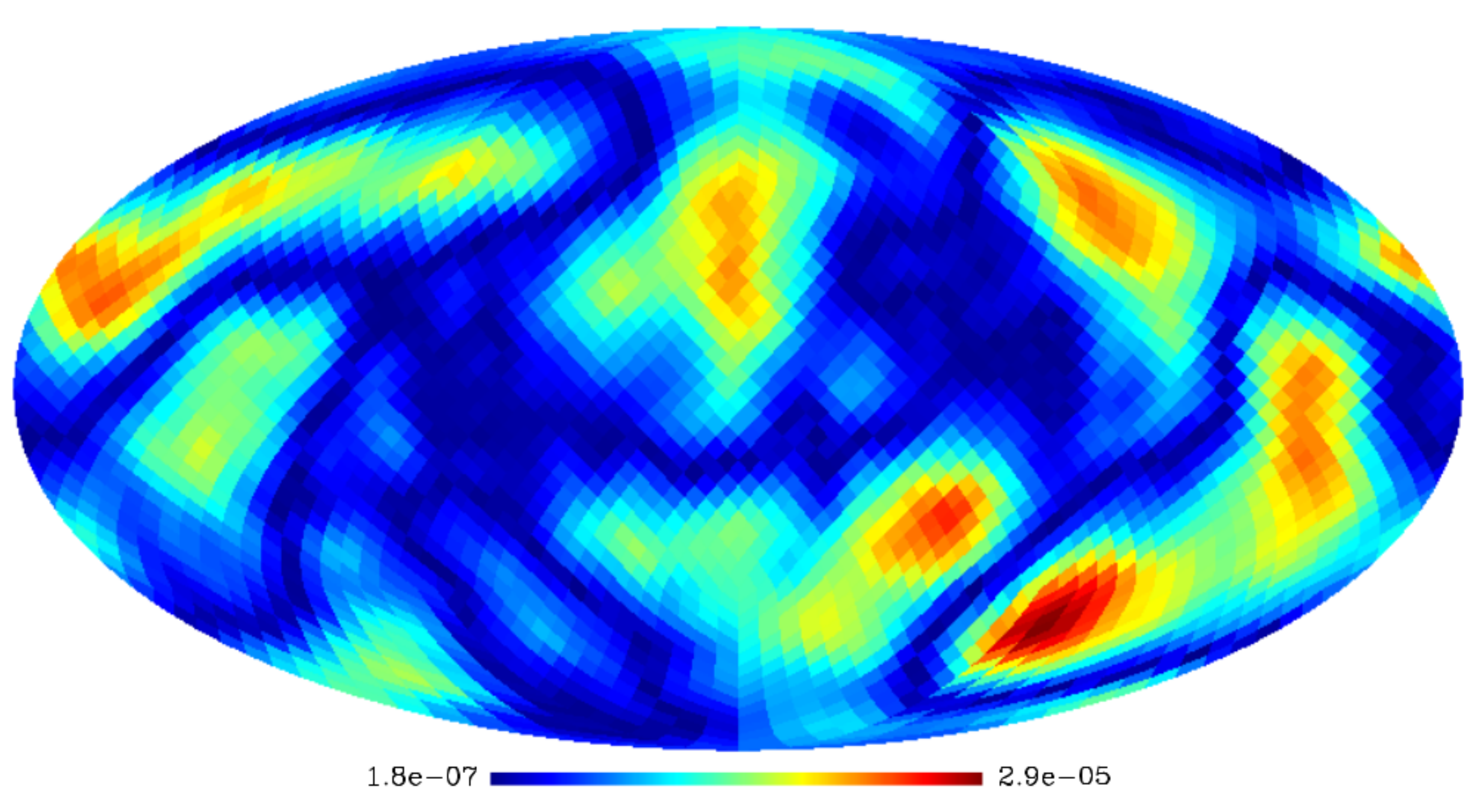}
		\caption{In the top panel we show the trace plot of Wiener filter, $F_i$ values  estimated 
                from the cleaned map obtained at each iteration for a randomly chosen Gibbs chain and 
                 for chosen $500$ samples. We  see clearly that the the values of $F_i$ are very close to unity 
                  which implies that negligible noise levels are present in the cleaned E mode maps. In 
                the bottom panel we show the difference 
                    between the standard deviation maps obtained
		using the Wiener filtered cleaned E mode map and the set of cleaned E mode  maps 
              obtained without any Wiener filtering. The  difference between both these standard deviation maps
              is very small indicating negligible detector noise level is present in the cleaned E maps.}
		\label{weiner}
	\end{figure}
	In the current analysis our simulated input CMB maps have realistic level of detector
	noise generated using the noise models corresponding to COrE polarization maps after 
        $4$ years of observations. Since we are working
	at pixel resolution $N_{\tt side} = 16$ and smoothed by a Gaussian beam of FWHM = $9^{\circ}$,
	at this large smoothing, the detector noise present in the smoothed CMB E mode maps is low. The
	comparison of detector noise power for the multipole range $\ell$ = 2 $-$ 32 for all
	COrE frequencies is shown in Fig.~\ref{noisecls} along with the CMB EE power spectrum,
	obtained after removing the beam and pixel effects, as brown line. Even though, we see
	that the noise power at all multipoles for all COrE frequencies is 
	lower than the CMB E mode power, it is important to study whether the presence of detector
	noise can lead to any residual noise bias in the best fit E mode cleaned map.
	
	In order to understand the level of residual noise present in the cleaned CMB E-mode maps obtained
	at each iteration in our method, we performed Wiener filtering of the corresponding
	cleaned map. We first estimated the $15 \times 15$ empirical noise covariance matrix $(C_{\tt N})$ in pixel space 
        by cross-correlating
	the simulated noise maps of all $15$ COrE frequencies. The noise covariance matrix   is then used to estimate
	the noise variance $(\sigma_{\tt N}^2)$ in the cleaned E  mode map at each Gibbs step, $i$ following, 
	\begin{equation}
	\sigma_{\tt N}^{2i} = {\bf W}_{i \tt T}{\bf C}_{\tt N}{\bf W}_{i} \, ,
	\end{equation}
	where, ${\bf W}_i$ denotes  the $n \times 1$ weight vector obtained at $i^{\tt th}$ Gibbs step.
	In order to obtain the CMB variance, we randomly generate 50000 different CMB realizations
	at $N_{\tt side} = 16$ and at Gaussian beam resolution of FWHM = $9^{\circ}$. The CMB
	variance ($\sigma_{\tt CMB}^2$) is calculated by taking the mean of variance of each
	individual CMB realizations. Finally, we estimate the Wiener filter, $F_i$ at each Gibbs
	iteration as follows
	\begin{equation}
	F_i = \frac{\sigma^2_{\tt CMB}}{\sigma^2_{\tt CMB} + \sigma^{2i}_{\tt N}} \, .
	\end{equation}
	We calculate the $F_i$ values of all the $99500$ cleaned maps obtained after burn-in rejection 
        from all the $10$ chains.  In the top panel of Fig.~\ref{weiner} we show the filter values
	corresponding to one chain for $500$ sample values. We see that the filter values are very close to
	unity indicating negligible residual noise in the cleaned maps. Filter from each iteration
	is then used to suppress noise from the corresponding cleaned CMB E mode map. In the bottom panel of Fig.~\ref{weiner} we
	show the difference between the standard deviation map obtained from  all such
	noise weighted cleaned maps and the standard deviation map from our 
	method which does not use any Wiener filtering. Since the difference of the two standard deviation maps are very small 
        we conclude that both the standard deviation maps matches closely
	indicating  no significant residual noise bias  is present in our cleaned E mode maps. Therefore noise bias in the
	best fit cleaned CMB map is also negligible.
	
	\section{Estimating Likelihood function  using Blackwell Rao estimator}
	\label{br}
	For cosmological parameter estimation, it is of utmost importance to be able
	to compute the likelihood of any arbitrary theoretical CMB EE angular power spectrum given
	the data. Although our method computes posterior density of  theoretical CMB E mode  angular power
	spectrum, it is in effect a discrete estimation of the underlying power spectrum. Interestingly,
	Blackwell-Rao~\cite{2005PhRvD..71j3002C} theorem can be used for  an improved  estimation 
        of the likelihood  of the CMB EE power spectrum consistent with the foreground removal algorithm 
        used in this article.
	The theorem says that 
	we can always find an estimator with an equal or better efficiency than the initial estimator 
        by taking its conditional expectation with respect to a  sufficient statistic
	 The transformed  estimator using Blackwell-Rao theorem is called Blackwell-Rao estimator. In this 
        section we calculate the Blackwell-Rao estimates of likelihood function of  theoretical CMB E mode angular
	power spectrum by using $99500$ samples of theoretical angular power spectrum from our method. The 
       Blackwell-Rao expression for likelihood function of  $C_\ell^E$ given
	$\hat C_{\ell}^{E,i}$  is given by
	\begin{equation}
	P(C_{\ell}^E|\hat C_{\ell}^{E,i})  \propto  \frac{1}{\hat C_{\ell}^{E,i}}
	\left(\frac{\hat C_{\ell}^{E,i}}{C_{\ell}^E}\right)^{{(2\ell+1)}/{2}}
	\, \textrm{exp}\left[{-\frac{2\ell+1}{2} \frac{\hat C_{\ell}^{E,i}}{C_{\ell}^E}}\right]\, ,
	\label{eq:gauss_lnL}
	\end{equation}
	where $\hat C_{\ell}^{E,i}$ is the $i^{th}$ realization of the power spectrum obtained after burn-in
        rejection from all Gibbs chains.
	The likelihood function for $C_\ell^E$ given data $\textbf{D}$ now can be approximated  as~\cite{2005PhRvD..71j3002C}
	\begin{equation}
	P(C_{\ell}^E|\textbf{D}) \approx \frac{1}{N}\sum_{i=1}^{N}
	P(C_{\ell}^E|\hat C_{\ell}^{E,i}) \, ,
	\end{equation}
	where $N$ denotes total the number of Gibbs samples from all chains  after burn-in rejection.

	\begin{figure*}
		\hspace{-1cm}
		\includegraphics[scale=0.7]{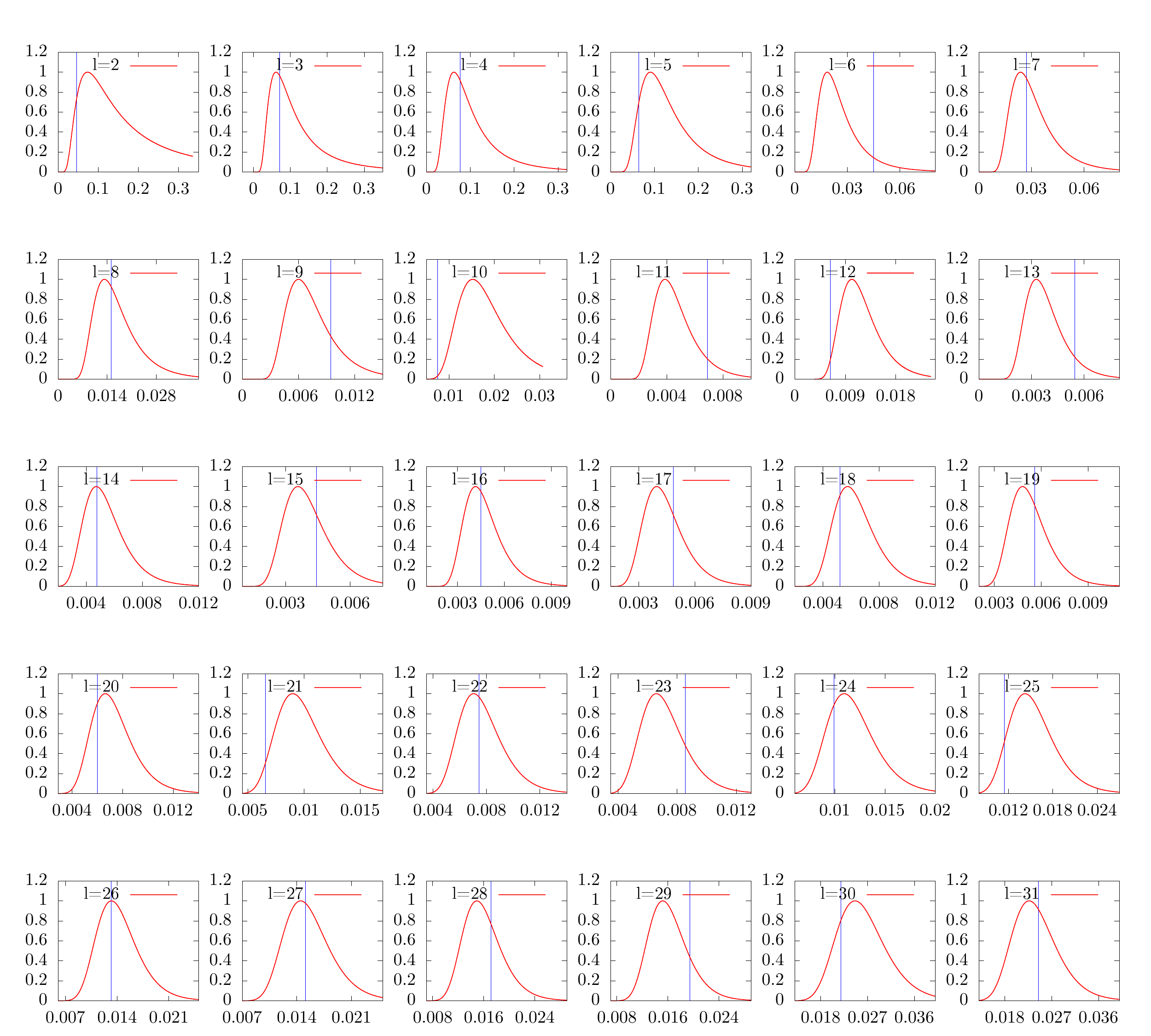}
		\caption{Plot showing the Blackwell-Rao estimates of the likelihood functions for the multipoles 
                 $2 \leq \ell \leq 31$. The horizontal axis is plotted 
               as  $\ell(\ell +1)C_\ell^E/2\pi $ in the unit of $\mu K^2 $. The vertical lines correspond to the 
                   positions of the CMB theoretical E mode angular power spectrum used to  generate the 
                 random realization of input CMB E mode map used in this work.}
		\label{BR}
	\end{figure*}

\begin{figure*}
       \includegraphics[scale=0.73]{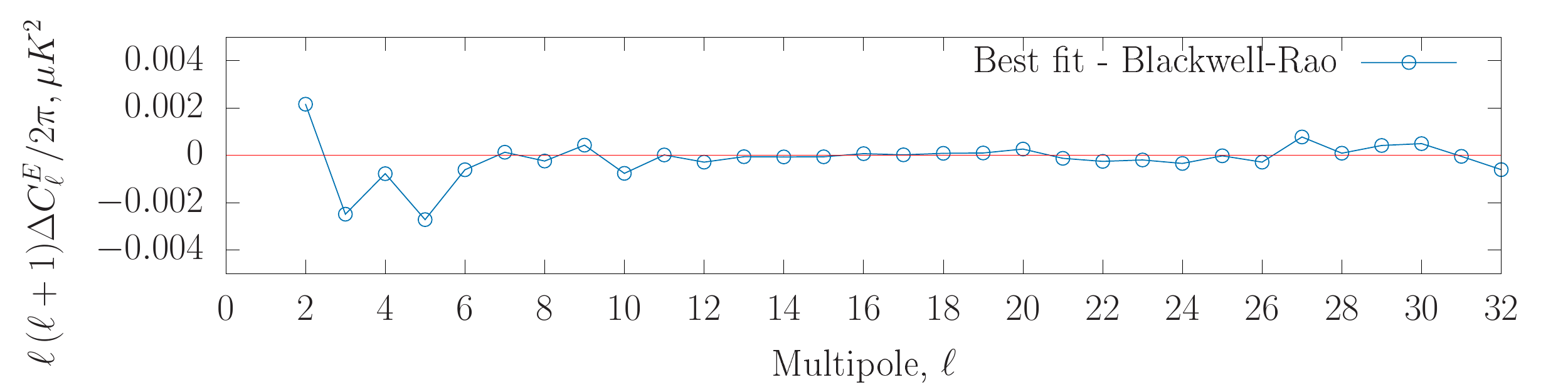}
        \caption{Figure showing the difference between the best-fit $C_\ell^E$ and the peak locations of the  
       E mode theoretical angular power spectrum obtained from the Blackwell-Rao likelihood functions. Both these 
       estimates match very well with each other.  }
        \label{bf-BR}
       \end{figure*}

	We show the results in Fig.~\ref{BR}. At low multipoles the likelihood  distribution is
	highly asymmetric with a long tail and as we go to higher multipoles
	the distribution becomes more symmetric. The  positions corresponding to the theoretical power spectrum 
        that was used to generate random realization of CMB E mode map used in this analysis is also shown by blue 
        vertical lines for comparison. The peak positions of likelihood functions of this plot match very well 
        with the best-fit spectrum which was shown in Fig.~\ref{cleanedclfig}. We show the difference between the two in Fig.~\ref{bf-BR}. 
        {\it An interesting property of the estimated likelihood 
        functions is that they are  entirely independent on the explicit foreground models. Hence the likelihood 
        functions are not affected by any modeling uncertainty of the polarized foregrounds. Moreover, these likelihood 
        functions are not affected  by the residual E mode foregrounds since we use a large number of input 
         frequency maps which results in a negligible foreground residual effects  in the cleaned E mode map
        and its theoretical angular power spectrum.}
         In a future work these  estimators will be  directly integrated
	 in cosmological parameter estimation using CMB EE angular power spectrum over large angular scales.
	
	\section{Testing convergence of the method}
	\label{convergence}
	Our implementation  of the  algorithm consists of 10 chains with different
	starting points and each chain with 10000 Gibbs steps. Overall after rejecting 50
	initial samples of CMB cleaned maps and CMB theoretical angular power spectrum in each
	chain, we have a total of 99500 samples. It is important to check whether in each chain,
	a chain length of 10000 Gibbs steps the samples have converged in order to ascertain
	the sampled distribution converged to the targeted CMB joint posterior distribution.
	Convergence of the chains also ensures that the final samples no longer depends on the
	initial points. A  powerful diagnostic to check for the convergence is the Gelman Rubin
	statistic~\citep{gelman1992} which proposes that lack of convergence in a
	chain is better diagnosed in the presence of  parallel chains with different initial
	points. We discuss this method.
	
	Let us consider a simulation for estimating the posterior of a model with parameter
	$\phi$. Let there are  $N$ number of chains with  $L$ being the number of steps
	after rejecting the samples during the burn-in phase. Let us assume that the
	sample posterior mean and variance are given by $\bar{\phi}_n$  and $(\bar{\sigma_n})^2$
	respectively for $n^{th}$ chain using all the $L$ samples in the chain. Then the
	between-chain $(B/L)$ and within-chain variances $(W)$ are respectively given by
	\begin{eqnarray}
	{B} =\frac{L}{N-1} \sum_{n=1}^{N}(\bar{\phi}_n-\bar{\phi})^2
	\end{eqnarray}
	
	\begin{eqnarray}
	{W} = \frac{1}{N} \sum_{n=1}^{N}(\bar{\sigma}_n)^2
	\end{eqnarray}
	where $\bar{\phi}$ is the mean of the samples from all the sample chains,
$\bar{\phi} = \frac{1}{N}\Bigl(\sum_{n=1}^{N}{\bar \phi}_n\Bigr)$

 The  pooled posterior variance is defined as follows

	\begin{eqnarray}
	{{V}} = \frac{L-1}{L}W  + \frac{N+1}{NL}B
	\end{eqnarray}
	The Gelman-Rubin statistic $R$ is then given by
	
	\begin{eqnarray}
	{R} = \sqrt\frac{{V}}{W}
	\end{eqnarray}
	According to~\cite{gelman1992}, the chain converges when the value of R approaches
	unity ($L\rightarrow \infty$) which in turn  implies the sampled distribution is same  
	as the targeted distribution.
	
\begin{figure}
	
		\centering
		\includegraphics[scale=0.57]{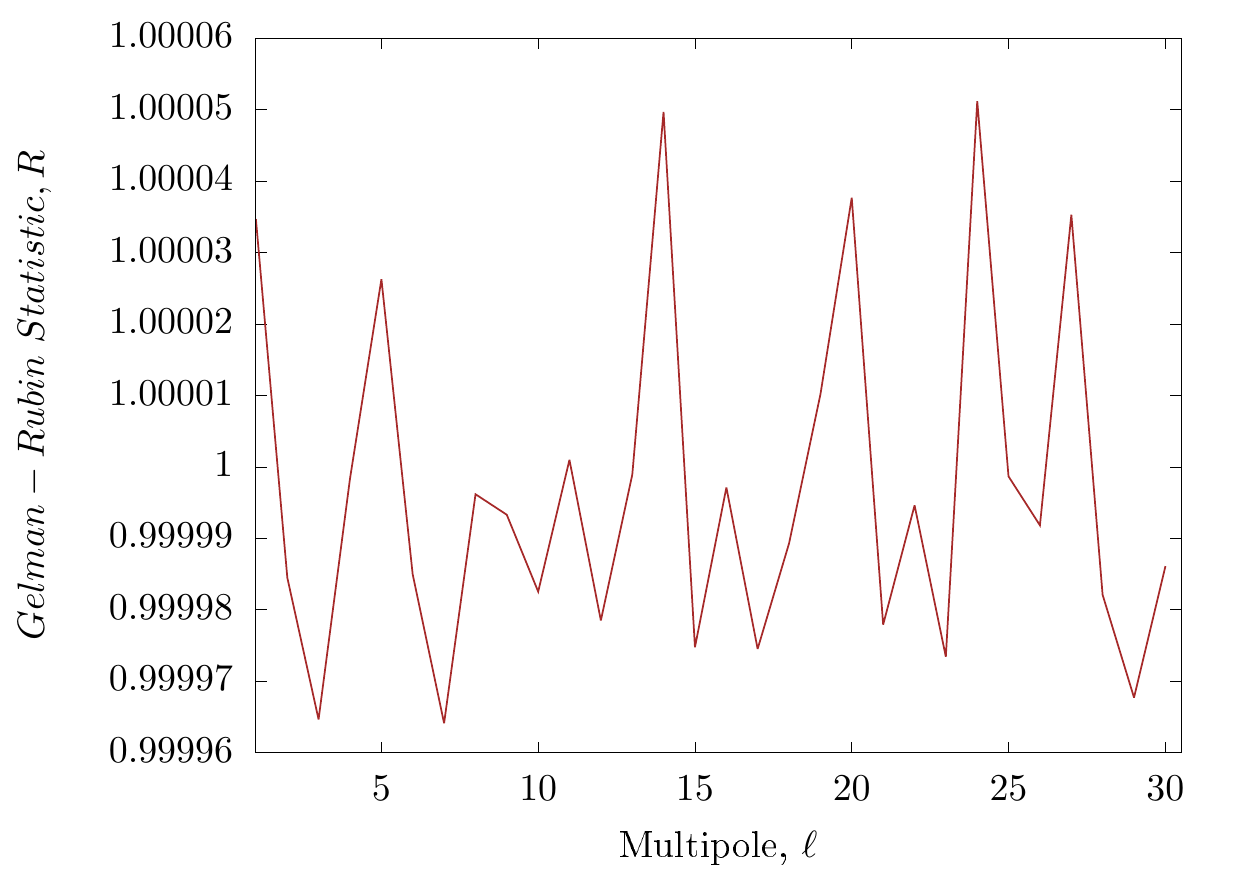}
		\includegraphics[scale=0.3]{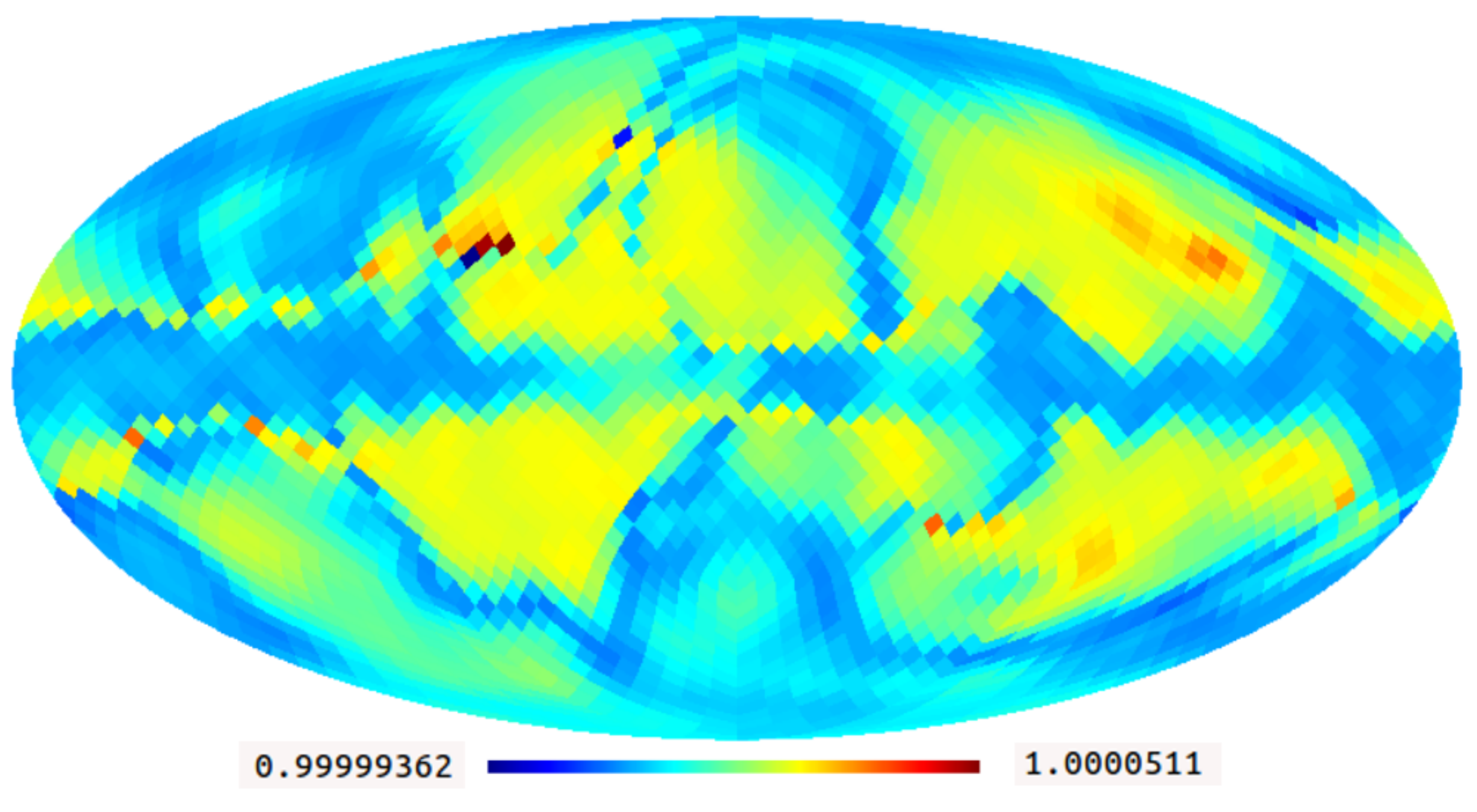}
		\caption{In the top panel we show the Gelman Rubin statistic  R value for each
			multipole. The bottom panel shows the same for the pixel values. From the figure  its clear  
			that the R values are close to 1 implying the chains have achieved convergence.}
		\label{gelmanfig}
	\end{figure}

	In the top panel of Fig.~\ref{gelmanfig}, we shown the Gelman-Rubin statistic, R, for
	the CMB theoretical angular power spectrum samples at all multipoles. The value of R
	lies well within 0.99996 and 1.00004 implying convergence. In the bottom panel of
	Fig.~\ref{gelmanfig}, we show the map of R values for all pixels. We see that
	the R value lies with in 0.99999362 and 1.0000511 for all the pixels implying again
	that Gibbs chains have converged.
	\section{Discussion and Conclusion}
	\label{disc}

         Accurate measurement of CMB E mode signal provide valuable physical information about the 
        ionization history of the universe. In our work,  we estimate CMB E mode signal and its theoretical angular power spectrum joint 
        posterior density over large angular scales of the sky using  simulated observations of $15$ 
        frequency maps between $60$ to $340$ GHz of future generation COrE satellite observations. We 
        employ the ILC method augmented by Gibbs sampling technique to draw samples from the joint density. 
        The input frequency maps contain E mode foregrounds and detector 
        noise contamination along with CMB E mode signal compatible to  Planck 2015 theoretical angular power 
        spectrum~\cite{2016A&A...594A..13P}. Foregrounds in input maps 
        consist of two major polarized components, namely,  the synchrotron and thermal dust emissions. 
        The detector noise level used in this work is compatible to $4$ years of observations of COrE. 
        We run a total of $10$ Gibbs chains each comprising  $10000$ numbers. We reject $50$ samples 
        from each chain due to burn-in removal which results in  a total of $99500$ usable samples. We 
        test convergence of these chains by evaluating Gelman-Rubin statistic and conclude reliable 
        convergence has been achieved in these chains. 

        Using the complete joint posterior density we evaluate the marginal densities of the CMB E mode 
        map and theoretical angular power spectrum. We see that the best-fit cleaned  CMB E map agree 
        excellently with the input CMB E mode map. The best-fit E mode angular power spectrum agree with 
        the sky angular power spectrum very well. The error in reconstruction is CMB E mode map is small
        for almost all over the sky. The reconstruction error is seen to be large only at  a couple of 
        pixels at the galactic center with a standard deviation of $\sim 0.0061$ $\mu K$. The samples of 
        theoretical E mode angular power spectra are used to estimate the asymmetric error levels on the corresponding 
        best-fit estimates.  The best-fit E mode angular power spectrum does not show any indication of presence 
        of a bias in the multipole range $2 \le \ell \le 32$ studied in this work.

         Since polarization signal is weak, presence of significant detector  noise  in the  input frequency maps 
        may lead to foreground residuals in the cleaned maps as well as foreground residuals and detector noise
        bias in the cleaned E mode angular power spectrum. In order to asses the level of detector noise in the cleaned 
        maps obtained in the Gibbs chains  we perform Wiener filtering of all the sampled clean CMB
        E mode maps obtained. We find that the Wiener filtered CMB E maps agree  very well 
        with the unfiltered maps implying detector noise can safely be ignored for the large angular scales 
        analysis for CMB E mode using $4$ year COrE  noise levels.

        Using all  Gibbs samples of CMB theoretical E mode  angular power spectrum after burn in rejection
       we estimate the likelihood function of any arbitrary angular power spectrum given the simulated 
        data of $15$ COrE frequency maps using Blackwell-Rao estimator.  The likelihood function can be directly integrated 
        to a Markov Chain Monte Carlo algorithm to estimate the cosmological parameters relevant 
        for large angular scales CMB E mode observations. {\it The CMB E mode component reconstruction method 
        of this article thus possesses an attractive property of accurately and reliably used in 
        cosmological parameter estimation methodology by predicting exact likelihood function of the E mode 
        theoretical angular power spectrum given the observed data after performing E mode
        foreground removal.}

        {\it  The method outlined in this work has another excellent advantage that the estimated  CMB  joint 
        posterior is not affected  by the inaccuracies in foreground modelling since the weight factors 
        for linear combination of input maps are estimated without assuming any foreground model.} This 
        is particularly very advantageous since the CMB E mode signal is very weak compared to the E mode
         foregrounds. A small modelling uncertainty in the foregrounds if exists may lead to large error (or 
        even a bias) in the reconstructed CMB E mode signal, if explicit foreground modelling is necessary 
        in order to remove the foregrounds. Moreover, the  effects of foreground residuals in the  reconstructed 
        joint posterior density following our method is negligible since we have 
        a large number of input frequency maps  which is larger than typical number of independent spectral 
        index parameters required for modelling the polarized foregrounds.   In a future article, we will extend the
        work for reconstruction of CMB B mode over large angular scales.

	This work is based on the CMB observations of Planck an ESA mission.
	For our work we have used the freely publicly available~\cite{2005ApJ...622..759G}
	HEALPix  Software package~\url{http://healpix.sourceforge.net}
	for some of the analysis of this work.We acknowledge the use of Planck
	Legacy Archive (PLA) and the Legacy Archive for Microwave Background Data
	Analysis (LAMBDA). LAMBDA is a part of the High Energy Astrophysics Science
	Archive Center (HEASARC). HEASARC/LAMBDA is supported by the Astrophysics
	Science Division at the NASA Goddard Space Flight Center.

	
	
	

\begin{thebibliography}{}
		\expandafter\ifx\csname natexlab\endcsname\relax\def\natexlab#1{#1}\fi
		
		\bibitem[{{Baccigalupi}(2003)}]{2003NewAR..47.1127B}
		{Baccigalupi}, C. 2003, New Astron. Rev., 47, 1127
		
		\bibitem[{Basak \& Delabrouille(2011)}]{10.1111/j.1365-2966.2011.19770.x}
		Basak, S., \& Delabrouille, J. 2011, Mon.\ Not.\ Roy.\ Astron.\ Soc.\, 419, 1163
		
		\bibitem[{{Bennett} {et~al.}(2003){Bennett}, {Halpern}, {Hinshaw}, {Jarosik},
			{Kogut}, {Limon}, {Meyer}, {Page}, {Spergel}, {Tucker}, {Wollack}, {Wright},
			{Barnes}, {Greason}, {Hill}, {Komatsu}, {Nolta}, {Odegard}, {Peiris},
			{Verde}, \& {Weiland}}]{Bennet2003}
		{Bennett}, C.~L., {Halpern}, M., {Hinshaw}, G., {et~al.} 2003, The
		Astrophysical Journal Supplement, 148, 1
		
		\bibitem[{{Bottino} {et~al.}(2010){Bottino}, {Banday}, \&
			{Maino}}]{2010MNRAS.402..207B}
		{Bottino}, M., {Banday}, A.~J., \& {Maino}, D. 2010, Mon.\ Not.\ Roy.\ Astron.\ Soc.\, 402, 207
		
		
		
		
		\bibitem[{{Bouchet} {et~al.}(1999){Bouchet}, {Prunet}, \&
			{Sethi}}]{1999MNRAS.302..663B}
		{Bouchet}, F.~R., {Prunet}, S., \& {Sethi}, S.~K. 1999, Mon.\ Not.\ Roy.\ Astron.\ Soc.\, 302, 663
		
		\bibitem[{{Bunn} {et~al.}(1994){Bunn}, {Fisher}, {Hoffman}, {Lahav}, {Silk}, \&
			{Zaroubi}}]{1994ApJ...432L..75B}
		{Bunn}, E.~F., {Fisher}, K.~B., {Hoffman}, Y., {et~al.} 1994, The Astrophysical
		Journal Letters, 432, L75
		
		\bibitem[{{Carretti}(2010)}]{2010ASPC..438..276C}
		{Carretti}, E. 2010, in Astronomical Society of the Pacific Conference Series,
		Vol. 438, The Dynamic Interstellar Medium: A Celebration of the Canadian
		Galactic Plane Survey, ed. R.~{Kothes}, T.~L. {Landecker}, \& A.~G. {Willis},
		276
		
		\bibitem[{{Delabrouille} {et~al.}(2009){Delabrouille}, {Cardoso}, {Le Jeune},
			{Betoule}, {Fay}, \& {Guilloux}}]{Delabrouille2009}
		{Delabrouille}, J., {Cardoso}, J.-F., {Le Jeune}, M., {et~al.} 2009, Astronomy
		\& Astrophysics, 493, 835
		
		\bibitem[{{Dickinson}(2016)}]{2016arXiv160603606D}
		{Dickinson}, C. 2016, arXiv e-prints, arXiv:1606.03606
		
		\bibitem[{Eisenstein {et~al.}(1999)Eisenstein, Hu, \&
			Tegmark}]{Eisenstein_1999}
		Eisenstein, D.~J., Hu, W., \& Tegmark, M. 1999, The Astrophysical Journal, 518,
		2
		
		\bibitem[{{Eriksen} {et~al.}(2008){Eriksen}, {Jewell}, {Dickinson}, {Banday},
			{G{\'o}rski}, \& {Lawrence}}]{Eriksen2008}
		{Eriksen}, H.~K., {Jewell}, J.~B., {Dickinson}, C., {et~al.} 2008, The
		Astrophysical Journal, 676, 10
		
		\bibitem[{Gelman \& Rubin(1992)}]{gelman1992}
		Gelman, A., \& Rubin, D.~B. 1992, Statist. Sci., 7, 457
		
		\bibitem[{{Gold} {et~al.}(2009){Gold}, {Bennett}, {Hill}, {Hinshaw}, {Odegard},
			{Page}, {Spergel}, {Weiland}, {Dunkley}, {Halpern}, {Jarosik}, {Kogut},
			{Komatsu}, {Larson}, {Meyer}, {Nolta}, {Wollack}, \& {Wright}}]{Gold2009}
		{Gold}, B., {Bennett}, C.~L., {Hill}, R.~S., {et~al.} 2009, The Astrophysical
		Journal Supplement, 180, 265
		
		\bibitem{Challinor:2004pr} 
		  A.~Challinor and G.~Chon,
		  Mon.\ Not.\ Roy.\ Astron.\ Soc.\  {\bf 360}, 509 (2005)
		  [astro-ph/0410097].
		
		\bibitem[{{G{\'o}rski} {et~al.}(2005){G{\'o}rski}, {Hivon}, {Banday}, {Wand
				elt}, {Hansen}, {Reinecke}, \& {Bartelmann}}]{2005ApJ...622..759G}
		{G{\'o}rski}, K.~M., {Hivon}, E., {Banday}, A.~J., {et~al.} 2005, The
		Astrophysical Journal, 622, 759
		
		\bibitem[{{Grasso} \& {Rubinstein}(2001)}]{Grasso2001}
		{Grasso}, D., \& {Rubinstein}, H.~R. 2001, Physics Reports, 348, 163
		
		\bibitem[{Hadzhiyska \& Spergel(2019)}]{PhysRevD.99.043537}
		Hadzhiyska, B., \& Spergel, D. 2019, Phys. Rev. D, 99, 043537
		
		\bibitem[{{Hu} \& {White}(1997)}]{Hu1997}
		{Hu}, W., \& {White}, M. 1997, arxiv, 2, 323
		
		\bibitem[{Ichiki(2014)}]{10.1093/ptep/ptu065}
		Ichiki, K. 2014, Progress of Theoretical and Experimental Physics, 2014,
		
		
		\bibitem[{Kaplinghat {et~al.}(2003)Kaplinghat, Knox, \&
			Song}]{PhysRevLett.91.241301}
		Kaplinghat, M., Knox, L., \& Song, Y.-S. 2003, Phys. Rev. Lett., 91, 241301
		
		\bibitem[Zaldarriaga \& Seljak(1997)]{1997PhRvD..55.1830Z} Zaldarriaga, M., \& Seljak, U.\ 1997, Phys.Rev.D, 55, 1830
		
		
		\bibitem[{Keating {et~al.}(1998)Keating, Timbie, Polnarev, \&
			Steinberger}]{Keating_1998}
		Keating, B., Timbie, P., Polnarev, A., \& Steinberger, J. 1998, The
		Astrophysical Journal, 495, 580
		
		\bibitem[{{Kim} {et~al.}(2008){Kim}, {Naselsky}, \& {Christensen}}]{Kim2008}
		{Kim}, J., {Naselsky}, P., \& {Christensen}, P.~R. 2008, Physical Review D, 77,
		103002
		
		\bibitem[{{Leach} {et~al.}(2008){Leach}, {Cardoso}, {Baccigalupi}, {Barreiro},
			{Betoule}, {Bobin}, {Bonaldi}, {Delabrouille}, {de Zotti}, {Dickinson},
			{Eriksen}, {Gonz{\'a}lez-Nuevo}, {Hansen}, {Herranz}, {Le Jeune},
			{L{\'o}pez-Caniego}, {Mart{\'{\i}}nez-Gonz{\'a}lez}, {Massardi}, {Melin},
			{Miville-Desch{\^e}nes}, {Patanchon}, {Prunet}, {Ricciardi}, {Salerno},
			{Sanz}, {Starck}, {Stivoli}, {Stolyarov}, {Stompor}, \& {Vielva}}]{Leach2008}
		{Leach}, S.~M., {Cardoso}, J.-F., {Baccigalupi}, C., {et~al.} 2008, Astronomy
		\& Astrophysics, 491, 597
		
		\bibitem[{{Maino} {et~al.}(2003){Maino}, {Banday}, {Baccigalupi}, {Perrotta},
			\& {G{\'o}rski}}]{2003MNRAS.344..544M}
		{Maino}, D., {Banday}, A.~J., {Baccigalupi}, C., {Perrotta}, F., \&
		{G{\'o}rski}, K.~M. 2003, Mon.\ Not.\ Roy.\ Astron.\ Soc.\,
		344, 544
		
		\bibitem[{{Maino} {et~al.}(2002){Maino}, {Farusi}, {Baccigalupi}, {Perrotta},
			{Banday}, {Bedini}, {Burigana}, {De Zotti}, {G{\'o}rski}, \&
			{Salerno}}]{2002MNRAS.334...53M}
		{Maino}, D., {Farusi}, A., {Baccigalupi}, C., {et~al.} 2002, Mon.\ Not.\ Roy.\ Astron.\ Soc.\, 334, 53
		
		\bibitem[{{Padmanabhan} \& {Finkbeiner}(2005)}]{Padmanabhan2005}
		{Padmanabhan}, N., \& {Finkbeiner}, D.~P. 2005, Physical Review D, 72, 023508
		
		\bibitem[{{Penzias} \& {Wilson}(1965)}]{Penzias1965}
		{Penzias}, A.~A., \& {Wilson}, R.~W. 1965, The Astrophysical Journal, 142, 419
		
		\bibitem[{{Purkayastha} \& {Saha}(2017)}]{2017arXiv170702008P}
		{Purkayastha}, U., \& {Saha}, R. 2017, arXiv e-prints, arXiv:1707.02008
		
		\bibitem[{{Remazeilles} {et~al.}(2018){Remazeilles}, {Banday}, {Baccigalupi},
			{Basak}, {Bonaldi}, {De Zotti}, {Delabrouille}, {Dickinson}, {Eriksen},
			{Errard}, {Fernandez-Cobos}, {Fuskeland}, {Herv{\'\i}as-Caimapo},
			{L{\'o}pez-Caniego}, {Martinez-Gonz{\'a}lez}, {Roman}, {Vielva}, {Wehus},
			{Achucarro}, {Ade}, {Allison}, {Ashdown}, {Ballardini}, {Banerji},
			{Bartlett}, {Bartolo}, {Baumann}, {Bersanelli}, {Bonato}, {Borrill},
			{Bouchet}, {Boulanger}, {Brinckmann}, {Bucher}, {Burigana}, {Buzzelli},
			{Cai}, {Calvo}, {Carvalho}, {Castellano}, {Challinor}, {Chluba}, {Clesse},
			{Colantoni}, {Coppolecchia}, {Crook}, {D'Alessandro}, {de Bernardis}, {de
				Gasperis}, {Diego}, {Di Valentino}, {Feeney}, {Ferraro}, {Finelli},
			{Forastieri}, {Galli}, {Genova-Santos}, {Gerbino}, {Gonz{\'a}lez-Nuevo},
			{Grandis}, {Greenslade}, {Hagstotz}, {Hanany}, {Handley},
			{Hernandez-Monteagudo}, {Hills}, {Hivon}, {Kiiveri}, {Kisner}, {Kitching},
			{Kunz}, {Kurki-Suonio}, {Lamagna}, {Lasenby}, {Lattanzi}, {Lesgourgues},
			{Lewis}, {Liguori}, {Lindholm}, {Luzzi}, {Maffei}, {Martins}, {Masi},
			{Matarrese}, {McCarthy}, {Melin}, {Melchiorri}, {Molinari}, {Monfardini},
			{Natoli}, {Negrello}, {Notari}, {Paiella}, {Paoletti}, {Patanchon}, {Piat},
			{Pisano}, {Polastri}, {Polenta}, {Pollo}, {Poulin}, {Quartin},
			{Rubino-Martin}, {Salvati}, {Tartari}, {Tomasi}, {Tramonte}, {Trappe},
			{Trombetti}, {Tucker}, {Valiviita}, {Van de Weijgaert}, {van Tent}, {Vennin},
			{Vittorio}, {Young}, \& {Zannoni}}]{2018JCAP...04..023R}
		{Remazeilles}, M., {Banday}, A.~J., {Baccigalupi}, C., {et~al.} 2018, Journal
		of Cosmology and Astroparticle Physics, 2018, 023
		
		\bibitem[{Rocha {et~al.}(2004)Rocha, Trotta, Martins, Melchiorri, Avelino,
			Bean, \& Viana}]{10.1111/j.1365-2966.2004.07832.x}
		Rocha, G., Trotta, R., Martins, C. J. A.~P., {et~al.} 2004, Mon.\ Not.\ Roy.\ Astron.\ Soc.\, 352, 20
		
		\bibitem[{{Saha}(2011)}]{Saha2011}
		{Saha}, R. 2011, The Astrophysical Journal Letters, 739, L56
		
		\bibitem[{{Saha} {et~al.}(2008){Saha}, {Prunet}, {Jain}, \&
			{Souradeep}}]{Saha2008}
		{Saha}, R., {Prunet}, S., {Jain}, P., \& {Souradeep}, T. 2008, Physical Review
		D, 78, 023003
		
		\bibitem[{{Samal} {et~al.}(2010){Samal}, {Saha}, {Delabrouille}, {Prunet},
			{Jain}, \& {Souradeep}}]{Samal2010}
		{Samal}, P.~K., {Saha}, R., {Delabrouille}, J., {et~al.} 2010, The
		Astrophysical Journal, 714, 840
		
		
		\bibitem[{{Souradeep} {et~al.}(2006){Souradeep}, {Saha}, \& {Jain}}]{Saha2006}
		{Souradeep}, T., {Saha}, R., \& {Jain}, P. 2006, New Astronomy Reviews, 50, 854
		
		\bibitem[{{Stark}(1981)}]{Stark1981_b}
		{Stark}, R.~F. 1981, Mon.\ Not.\ Roy.\ Astron.\ Soc.\, 195,
		127
		
		\bibitem[{{Sudevan} \& {Saha}(2018{\natexlab{a}})}]{2018ApJ...867...74S}
		{Sudevan}, V., \& {Saha}, R. 2018{\natexlab{a}}, The Astrophysical Journal,
		867, 74
		
		\bibitem[{{Remazeilles} {et~al.}(2018){Remazeilles}, {Dickinson}, {Eriksen}, \&
		  {Wehus}}]{Rema2018}
		{Remazeilles}, M., {Dickinson}, C., {Eriksen}, H.~K., \& {Wehus}, I.~K. 2018,
		  MNRAS, 474, 3889


                 \bibitem[{{Planck Collaboration} {et~al.}(2016{\natexlab{f}}){Planck
 Collaboration}, {Aghanim}, {Ashdown}, {Aumont}, {Baccigalupi}, {Ballardini},
 {Banday}, {Barreiro}, {Bartolo}, {Basak}, {Benabed}, {Bernard}, {Bersanelli},
 {Bielewicz}, {Bonavera}, {Bond}, {Borrill}, {Bouchet}, {Boulanger},
 {Burigana}, {Calabrese}, {Cardoso}, {Carron}, {Chiang}, {Colombo}, {Comis},
 {Couchot}, {Coulais}, {Crill}, {Curto}, {Cuttaia}, {de Bernardis}, {de
 Zotti}, {Delabrouille}, {Di Valentino}, {Dickinson}, {Diego}, {Dor{\'e}},
 {Douspis}, {Ducout}, {Dupac}, {Dusini}, {Elsner}, {En{\ss}lin}, {Eriksen},
 {Falgarone}, {Fantaye}, {Finelli}, {Forastieri}, {Frailis}, {Fraisse},
 {Franceschi}, {Frolov}, {Galeotta}, {Galli}, {Ganga}, {G{\'e}nova-Santos},
 {Gerbino}, {Ghosh}, {Giraud-H{\'e}raud}, {Gonz{\'a}lez-Nuevo}, {G{\'o}rski},
 {Gruppuso}, {Gudmundsson}, {Hansen}, {Helou}, {Henrot-Versill{\'e}},
 {Herranz}, {Hivon}, {Huang}, {Jaffe}, {Jones}, {Keih{\"a}nen}, {Keskitalo},
 {Kiiveri}, {Kisner}, {Krachmalnicoff}, {Kunz}, {Kurki-Suonio}, {Lamarre},
 {Langer}, {Lasenby}, {Lattanzi}, {Lawrence}, {Le Jeune}, {Levrier}, {Lilje},
 {Lilley}, {Lindholm}, {L{\'o}pez-Caniego}, {Ma}, {Mac{\'{\i}}as-P{\'e}rez},
 {Maggio}, {Maino}, {Mandolesi}, {Mangilli}, {Maris}, {Martin},
 {Mart{\'{\i}}nez-Gonz{\'a}lez}, {Matarrese}, {Mauri}, {McEwen}, {Melchiorri},
 {Mennella}, {Migliaccio}, {Miville-Desch{\^e}nes}, {Molinari}, {Moneti},
 {Montier}, {Morgante}, {Moss}, {Natoli}, {Oxborrow}, {Pagano}, {Paoletti},
 {Patanchon}, {Perdereau}, {Perotto}, {Pettorino}, {Piacentini},
 {Plaszczynski}, {Polastri}, {Polenta}, {Puget}, {Rachen}, {Racine},
 {Reinecke}, {Remazeilles}, {Renzi}, {Rocha}, {Rosset}, {Rossetti}, {Roudier},
 {Rubi{\~n}o-Mart{\'{\i}}n}, {Ruiz-Granados}, {Salvati}, {Sandri},
 {Savelainen}, {Scott}, {Sirignano}, {Sirri}, {Soler}, {Spencer}, {Suur-Uski},
 {Tauber}, {Tavagnacco}, {Tenti}, {Toffolatti}, {Tomasi}, {Tristram},
 {Trombetti}, {Valiviita}, {Van Tent}, {Vielva}, {Villa}, {Vittorio},
 {Wandelt}, {Wehus}, {Zacchei}, \& {Zonca}}]{ThDustGNILC2016}
{Planck Collaboration}, {Aghanim}, N., {Ashdown}, M., {et~al.}
 2016{\natexlab{f}}, Astronomy and Astrophysics, 596, A109

		  
		  \bibitem[{{Delabrouille} {et~al.}(2013){Delabrouille}, {Betoule}, {Melin},
		    {Miville-Desch{\^e}nes}, {Gonzalez-Nuevo}, {Le Jeune}, {Castex}, {de Zotti},
		    {Basak}, {Ashdown}, {Aumont}, {Baccigalupi}, {Banday}, {Bernard}, {Bouchet},
		    {Clements}, {da Silva}, {Dickinson}, {Dodu}, {Dolag}, {Elsner}, {Fauvet},
		    {Fa{\"y}}, {Giardino}, {Leach}, {Lesgourgues}, {Liguori},
		    {Mac{\'{\i}}as-P{\'e}rez}, {Massardi}, {Matarrese}, {Mazzotta}, {Montier},
		    {Mottet}, {Paladini}, {Partridge}, {Piffaretti}, {Prezeau}, {Prunet},
		    {Ricciardi}, {Roman}, {Schaefer}, \& {Toffolatti}}]{PSM2013}
		  {Delabrouille}, J., {Betoule}, M., {Melin}, J.-B., {et~al.} 2013,Astronomy \& astrophysics , 553,
		    A96
		
		\bibitem[{{Sudevan} \& {Saha}(2018{\natexlab{b}})}]{2018arXiv181008872S}
		---. 2018{\natexlab{b}}, arXiv e-prints, arXiv:1810.08872
		
		\bibitem[{{Tegmark} {et~al.}(2003){Tegmark}, {de Oliveira-Costa}, \&
			{Hamilton}}]{Tegmark2003}
		{Tegmark}, M., {de Oliveira-Costa}, A., \& {Hamilton}, A.~J. 2003, Physical
		Review D, 68, 123523
		
		\bibitem[{{Tegmark} \& {Efstathiou}(1996)}]{Tegmark1996}
		{Tegmark}, M., \& {Efstathiou}, G. 1996, Mon.\ Not.\ Roy.\ Astron.\ Soc.\, 281, 1297
		
		

		\bibitem[{Zaldarriaga \& Seljak(1997)}]{PhysRevD.55.1830}
		Zaldarriaga, M., \& Seljak, U. c.~v. 1997, Phys. Rev. D, 55, 1830
		
		\bibitem[{Geman \& Geman(1984)}]{Gibbs1984}
		Geman, S., \& Geman, D. 1984, IEEE Trans. Pattern Anal. Mach. Intell., 6, 721
		
		\bibitem[{{Zaldarriaga} {et~al.}(1997){Zaldarriaga}, {Spergel}, \&
			{Seljak}}]{1997ApJ...488....1Z}
		{Zaldarriaga}, M., {Spergel}, D.~N., \& {Seljak}, U. 1997, The Astrophysical
		Journal, 488, 1
		
		\bibitem{Hiramatsu:2018nfa}
		T.~Hiramatsu, E.~Komatsu, M.~Hazumi and M.~Sasaki,
		Physical Review D {\bf 97}, no. 12, 123511 (2018)
		[arXiv:1803.00176 [astro-ph.CO]].
		
		\bibitem{Groeneboom:2008fz}
		N.~E.~Groeneboom and H.~K.~Eriksen,
		Astrophys.\ J.\  {\bf 690}, 1807 (2009)
		[arXiv:0807.2242 [astro-ph]].
		
		
		\bibitem{Eriksen:2007mx}
		H.~K.~Eriksen, J.~B.~Jewell, C.~Dickinson, A.~J.~Banday, K.~M.~Gorski and C.~R.~Lawrence,
		Astrophys.\ J.\  {\bf 676}, 10 (2008)
		[arXiv:0709.1058 [astro-ph]].
		
		\bibitem{Eriksen:2007mp}
		H.~K.~Eriksen, C.~Dickinson, J.~B.~Jewell, A.~J.~Banday, K.~M.~Gorski and C.~R.~Lawrence,
		Astrophys.\ J.\  {\bf 672}, L87 (2008)
		[arXiv:0709.1037 [astro-ph]].
		
		\bibitem{Wandelt:2003uk}
		B.~D.~Wandelt, D.~L.~Larson and A.~Lakshminarayanan,
		Phys.\ Rev.\ D {\bf 70}, 083511 (2004)
		[astro-ph/0310080].
		
		\bibitem[Penrose(1955)]{1955PCPS...51..406P} Penrose, R.\ 1955, Proceedings of the Cambridge Philosophical Society, 51, 406
		
		\bibitem{FernandezCobos:2011bm}
		R.~Fernandez-Cobos, P.~Vielva, R.~B.~Barreiro and E.~Martinez-Gonzalez,
		Mon.\ Not.\ Roy.\ Astron.\ Soc.\  {\bf 420}, no. 3, 2162 (2012)
		[arXiv:1106.2016 [astro-ph.CO]].
		
		\bibitem{Mather:1993ij}
		J.~C.~Mather {\it et al.},
		Astrophys.\ J.\  {\bf 420}, 439 (1994).
		
		\bibitem[{Metropolis {et~al.}(1953)Metropolis, Rosenbluth, Rosenbluth, Teller,
		  \& Teller}]{Metropolis1953}
		Metropolis, N., Rosenbluth, A.~W., Rosenbluth, M.~N., Teller, A.~H., \& Teller,
		  E. 1953, The Journal of Chemical Physics, 21, 1087
		
		\bibitem{Gold:2008kp}
		B.~Gold {\it et al.} [WMAP Collaboration],
		Astrophys.\ J.\ Suppl.\  {\bf 180}, 265 (2009)
		[arXiv:0803.0715 [astro-ph]].
		
		\bibitem{Bedini:2004dd}
		L.~Bedini, D.~Herranz, E.~Salerno, C.~Baccigalupi, E.~E.~Kuruouglu and A.~Tonazzini,
		astro-ph/0407108.
		\bibitem{Purkayastha:2020snm}
		U.~Purkayastha, V.~Sudevan and R.~Saha,
		arXiv:2003.13570 [astro-ph.CO].
		
	\bibitem[Planck Collaboration et al.(2016)]{2016A&A...594A..13P} Planck Collaboration, Ade, P.~A.~R., Aghanim, N., et al.\ 2016, Astronomy and Astrophysics, 594, A13
	
	\bibitem[Delabrouille et al.(2018)]{2018JCAP...04..014D} Delabrouille, J., de Bernardis, P., Bouchet, F.~R., et al.\ 2018, Journal of Cosmology and Astroparticle Physics, 2018, 014
	
	\bibitem[Chu et al.(2005)]{2005PhRvD..71j3002C} Chu, M., Eriksen, H.~K., Knox, L., et al.\ 2005, Physical Review D, 71, 103002	
	
	\bibitem[Sudevan \& Saha(2020)]{2020arXiv200102849S} Sudevan, V., \& Saha, R.\ 2020, arXiv e-prints, arXiv:2001.02849
	

	\end{thebibliography}


\end{document}